\documentclass[aip,pof,preprint]{revtex4-1}
\usepackage{graphicx}
\usepackage{subfigure,rotating,amsmath,float,amssymb,amsbsy}
\usepackage{color}
\usepackage{subfigure}

\newsavebox{\astrutbox}
\sbox{\astrutbox}{\rule[-5pt]{0pt}{20pt}}

\newcommand\bx{\ensuremath{\mathbf{x}}}

\newcommand\bfor{\ensuremath{\mathbf{f}}}
\newcommand\bu{\ensuremath{\mathbf{u}}}

\newcommand\lpl{\ensuremath{\left(}}
\newcommand\rp{\ensuremath{\right)}}
\newcommand\bxpp{\ensuremath{ \lpl \bx \rp}}

\newcommand\rmmax{\ensuremath{\mathrm{max} }}

\newcommand\dgam{\ensuremath{\dot{\gamma}}}
\newcommand\dgamax{\ensuremath{\dgam_{\rmmax}}}
\newcommand\hid{\ensuremath{h_{\mathrm{ini}}}}
\newcommand\hld{\ensuremath{h_{\mathrm{lat}}}}
\newcommand\hi{\ensuremath{\hid/a}}
\newcommand\hl{\ensuremath{\hld/a}}
\newcommand\hcw{\ensuremath{h_{\mathrm{cw}}}}

\newcommand\caAom{\ensuremath{\lpl Ca, \omgnon \rp}}
\newcommand\us{\ensuremath{U_{\mathrm{slip}}/a\dgamax}}
\newcommand\usd{\ensuremath{U_{\mathrm{slip}}}}
\newcommand\ul{\ensuremath{U_{\mathrm{lat}}/a\dgamax}}
\newcommand\uld{\ensuremath{U_{\mathrm{lat}}}}
\newcommand\uml{\ensuremath{\bar{U}_{\mathrm{lat}}/a\dgamax}}

\newcommand\bsg{\ensuremath{\boldsymbol{\sigma}}}

\newcommand\nstsm{\ensuremath{\bar{\Delta N}}}
\newcommand\Ca{\ensuremath{\mathrm{Ca}}}

\newcommand\hD{\ensuremath{\hat{D}}}
\newcommand\hDpk{\ensuremath{\hD_{\mathrm{peak}}}}
\newcommand\phiv{\ensuremath{\phi_{\mathrm{v}}}}

\def \Gs {{G_{\mathrm{s}}}}

\def \Caopt {\Ca_{\mathrm{opt}}}
\def \DN {\Delta N}
\def \adh {a / h_{\mathrm{ini}}}

\def \ulat {U_{\mathrm{lat}}}
\def \omgnon {\omega/\dot{\gamma}_{\mathrm{max}}}

\begin{document}
 \title{The dynamics of a capsule in a wall-bounded oscillating shear flow}

\begin{abstract}
The motion of an initially spherical capsule in a
wall-bounded oscillating shear flow is investigated via an accelerated boundary integral
implementation. The neo-Hookean model is used as the constitutive law of the capsule membrane.
The maximum wall-normal migration is observed when the oscillation period of the imposed shear is of the order of the 
relaxation time of the elastic membrane; hence, the optimal capillary 
number scales with the
inverse of the oscillation frequency and the ratio agrees well with the 
theoretical prediction in the limit of high-frequency oscillation.
The migration
velocity decreases monotonically with the frequency of the applied shear and the capsule-wall distance. 
We report a significant correlation between the capsule lateral migration and the normal stress difference induced in the flow.
The periodic variation of the capsule deformation is roughly in phase with that of the migration velocity
and normal stress difference, with twice the frequency of the imposed 
shear.
The maximum deformation increases linearly with the membrane elasticity before reaching a plateau at higher 
capillary numbers when the deformation is limited by the time over which
shear is applied in the same direction and not by the 
membrane deformability.
The maximum membrane deformation scales 
as the distance to the wall to the power $1/3$ as observed for capsules and droplets in 
near-wall steady shear flows.

\end{abstract}

\author{LaiLai Zhu}\email{lailaizhu00@gmail.com}
\affiliation{Linn\'{e} Flow Center and SeRC, KTH Mechanics, S-100 44 
Stockholm, Sweden.}
\affiliation{Laboratory of Fluid 
Mechanics and Instabilities,
Station 9, EPFL, 1105 Lausanne, Switzerland}

\author{Jean Rabault}
\affiliation{Linn\'{e} Flow Center and SeRC, KTH Mechanics, S-100 44 
Stockholm, Sweden.}
\affiliation{\'{E}cole Polytechnique, 91128 Palaiseau Cedex, France}
\affiliation{Current address: Mechanics Division, Department of 
Mathematics, University of Oslo, 0316 Oslo, Norway}

\author{Luca Brandt}
\affiliation{Linn\'{e} Flow Center and SeRC, KTH Mechanics, S-100 44 
Stockholm, Sweden.}
\date{\today}

\maketitle
\section{Introduction}

The dynamics of a capsule in external flows have attracted enormous interest
due to their physiological and biological significance, leading to the discovery of
a variety of interesting and complex phenomena. Since the observation of the tank-treading (TT)
motion of a human red blood cell (RBC) in the viscometer
~\cite{fischer1978tank,fischer1978red_science}, 
real cells and their models 
have been examined in shear flow, one of the simplest
flows, both experimentally, theoretically and
numerically in order to understand the motions of
biological cells \cite{dbb1980_asym,egg1998large,
kraus1996fluid, ha2000electrohydrodynamic,
sui08tank,lac04sphe,le10bend,lebedev2007dynamics,ma2009numerical,cordasco2013orbital,
omori2012reorientation,zhao2012shear,koleva2012deformation,peng2013deformation}. The 
focus was on an initially spherical capsule that undergoes shear-induced TT motion, reaching a steady ellipsoidal 
shape~\cite{dbb1981time,poz95shear}, when the membrane
elements rotate along the stationary configuration like the caterpillar
driving of a tank. The undeformed shape of biological cells is, however,  seldom a sphere, rather 
geometrically anisotropic, the bi-concave shape of undeformed RBCs being a 
classic example. 
This non-sphericity introduces a shape memory for the
capsules and tends to orient them in a preferential
direction~\cite{ufert2011cap_review_shear}. 
As a consequence, the capsules tank-tread and oscillate around a preferred direction
as the shear rate is above a certain value~\cite{keller1982motion, 
skotheim2007red}; this oscillatory behavior is termed as the swinging
mode~\cite{abkarian2007swinging}.

Oscillating shear flows have been investigated 
to account for the physiological pulsation, and more complex and diverse capsule dynamics explored. 
Experiments~\cite{nakajima1990deformation} showed that
the RBCs deform more in the retarding phase than in the accelerating phase. Cells in harmonically
modulated shear were examined by theoretical approaches~\cite{kessler2009elastic} to reveal a
resonance behavior; a particular combination of the oscillation frequency and phase can
induce the tumbling motion of a capsule which would otherwise swing under steady shear
flow. Dupire {\it et al.}~\cite{dupire2010chaotic} theoretically produced disordered motions of the RBCs under
sinusoidally varying flows with physiologically relevant parameters.
A similar chaotic
behavior was also reproduced in the analytical investigation by Noguchi~\cite{noguchi2010dynamic} who
found multiple limit-cycle solutions of a model RBC subject to an oscillating
shear flow at high frequency; their tumbling or tank-treading motions 
were found
at high or low shear amplitudes. The recent numerical simulations in
Ref.~\onlinecite{zhao2011dynamics} reproduce the two swinging modes of an initially oblate capsule
by varying the frequency of the external shear; these authors also document  the
high sensitivity of the motion of the capsule to its initial orientation, indirectly confirming
the chaotic motion reported in Ref.~\onlinecite{dupire2010chaotic}.

These previous  studies mainly focused on the dynamics of capsules in an unbounded oscillatory shear
flow. In the micro-circulation, however, cells like the RBCs  move in strongly
confined oscillating flows, an example being the physiological process
called vasomotion. This corresponds to the spontaneous and rhythmic oscillations of
large arteries as well as microvessels \cite{haddock2005rhythmicity,
shimamura1999tension,nilsson2003vasomotion}. Vasomotion is independent of the common
physiological pulsations induced by the heart beat, innervation or
respiration~\cite{aalkjaer2005vasomotion}, but due to the constriction and
dilation of the smooth muscle. Its possible benefits include reducing the time-averaged hydraulic resistance to the
blood flow \cite{meyer2002reassessing} and the consequent enhancement of oxygen
transport \cite{secomb1989effects,goldman2001computational}. 
It is therefore important to understand the dynamics of deformable particles like biological
cells in a confined oscillating flow. 

In addition to the complexity introduced by the oscillation, the influence of  wall confinement  on
the cell motion is crucial and has been, thus far, mostly investigated in steady 
flows 
where
cells exhibit cross-streamline migrations in most cases.
F\aa hraeus and Lindqvist~\cite{faahraeus1931viscosity} 
first unraveled the biological importance of
confinement. 
They discovered that the RBCs tend to deform significantly and migrate towards the
center of a microvessel owing to the hydrodynamic lift resulting from the interaction between
the cellular deformation and the wall; 
this migration facilitates a lower flow resistance and a more effective
mass transport. 
Simulations using the front-tracking method have been conducted by Doddi 
{\it et al.}~\cite{bagchi08_migration}
to investigate the cross-stream lift of a capsule towards the centerline in a Poiseuille flow; these authors observed that the migration 
velocity 
scales as the distance to the wall to the power $1/3$ but varies non-linearly 
with the capillary number.
The recent numerical study by Pranay {\it et al.}~\cite{pranay2012depletion},
based on an accelerated boundary integral method as that 
used here, investigates the lateral migration of an individual and a suspension of initially spherical capsules in 
wall-bounded Newtonian and viscoelastic shear flows. These authors show that the migration of the capsule depends on 
the capillary number, and that it is attenuated by the addition of 
polymers.
Migration has also been reported for vesicles: in an unbounded Poiseuille 
flow~\cite{kaoui2008lateral,mi09_ves_poi}, the migration velocity depends on the flow curvature, vesicle deformability
and viscosity contrast of the fluid inside and outside the vesicle; in a wall-bounded shear, the migration velocity is 
proportional to the wall normal component of the particle stresslet and the presence of the wall delays its transition from 
the tank-treading motion to trembling and 
tumbling~\cite{zhao2011dynamicsvesicle}.

Recently, two groups 
\cite{singh2014lateral,nix2014lateral} have numerically examined the 
case of a near-wall single capsule transported in a Newtonian shear flow, using a front-tracking and a boundary 
integral 
method respectively. Singh {\it et al.}~\cite{singh2014lateral} document the power-law relations between the migration 
velocity, capsule deformation and the capsule-wall distance, as previously done analytically for a 
droplet. These authors also propose a semi-analytical theory identifying two competing mechanisms influencing the 
lateral migration:
the interfacial stresses and the viscosity ratio; this theory agrees well with the simulation results
for large capsule-wall distances. Nix {\it et al.}~\cite{nix2014lateral} pay more attention to the correlation between the 
deformation of a capsule and its migration. These authors find  that an asymmetric capsule deformation reduces the 
migration velocity and this effect is compensated by an increase in the capsule stresslet, in turn enhancing its 
migration. 

Very recently, Matsunaga {\it et 
al.~\cite{matsunaga2015deformation} investigated the motion of capsule in an 
unbounded oscillating shear flow, observing an overshoot phenomenon when the 
maximum deformation of capsule can be larger than its deformation in a
steady shear.} In this manuscript, we extend the work in 
Ref.~\onlinecite{singh2014lateral}, Ref.~\onlinecite{nix2014lateral} and 
Ref.~\onlinecite{matsunaga2015deformation} by considering a 
near-wall individual capsule in oscillating plane shear flow. We hence 
introduce one additional factor, i.e., the
unsteadiness of the background flow that is a typical feature of biological 
environments.
We aim to understand 
the interplay between the wall and the flow unsteadiness, explaining its influence
on the dynamics of a capsule. 

The paper is organized as follows. After the description of the problem setup, we give a brief introduction
to our computational framework based on an accelerated boundary integral method suited for general geometries. We
firstly report the trajectories of capsules migrating from the wall and then investigate the dependence of the lateral 
migration velocity on the capsule capillary number, the capsule-wall distance and the frequency of 
the oscillation of the background shear. We further analyze the correlation between the migration velocity and the 
normal 
stress difference that arises from the viscoelastic effects due to the deformable particles. Finally, the deformation 
of the capsule is examined; the phase difference between the deformation and the applied shear is measured to quantify 
the induced viscoelasticity. The paper ends with a summary of the main findings.

\section{Problem setup and numerical methods}
\subsection{Problem setup}
\begin{figure}
\centering
	\includegraphics[width=0.8\columnwidth]
	{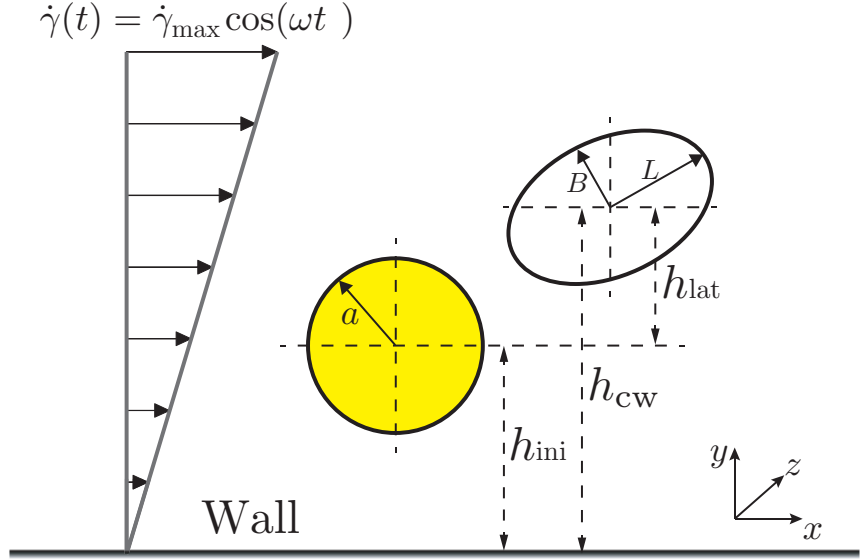}
\caption{(Color online) We consider a capsule whose undeformed shape is
a sphere of radius $a$, subject to an oscillating wall-bounded shear flow 
$\dgam \lpl t \rp =
\dgamax \cos \lpl \omega t \rp$. The center of mass of the capsule is initially 
 a distance  $\hid$ above
the wall and its lateral migration, measured from the initial position, is 
denoted $\hld$. The distance
between the capsule and the wall is thus $\hcw = \hid + \hld$.}
\label{fig:setup}
\end{figure}

We compute the motion and deformation of a capsule subject to a wall-bounded shear
flow, see the sketch in figure~\ref{fig:setup}. 
The undeformed shape of the capsule is a sphere with radius $a$ and  
center of mass initially located at a distance $\hi$  above the wall.
A time-periodic harmonic  shear flow $\dgam \lpl t \rp =
\dgamax \cos \lpl \omega t \rp$ is applied. $\dgamax$ denotes the maximum value of the shear,
and $\omega$ is the oscillation frequency. The characteristic time of the flow is thus $T=2\pi/\omega$.
In the current work, we study a capsule as a model cell and reproduce its 
motion and deformation in the micro-circulation or in micro-fluidic devices. We therefore assume that the capsule is
advected by a creeping flow and the Reynolds number $Re$, indicating the ratio of inertial over viscous forces, is zero.

The capsule is considered as a fluid-filled droplet enclosed by an infinitely thin elastic membrane.
The fluid inside and outside the capsule has the same density $\rho$ and the same dynamic 
viscosity $\mu$. For real biological cells, the viscosity of the fluid inside and outside is not necessarily 
the same; this is not taken into account here to limit the investigated parameter space.
The capsule is deformed by the fluid flow and the stress on its membrane is determined by the 
neo-Hookean constitutive law \cite{bar+wal+sal10}. The
membrane has a shear modulus $\Gs$ and zero bending stiffness. Elastic stresses
 develop on the surface of the capsule due to its deformation; this stress modifies the
surrounding flow in return. In this fluid-structure interaction problem, the viscous stress of the
flow and elastic stress of the deformable membrane compete with each other, their ratio defined
as the capillary number $\Ca=\mu\dgamax/G_{\mathrm{s}}$.

\subsection{Numerical method}
The membrane of the capsule is discretized into $N$ Lagrangian points. 
Neglecting the inertia
of the capsule, the elastic force balances the flow viscous force on its 
surface, 
\begin{equation}\label{eq:force_balance_point}
 \boldsymbol{\rho}^{\mathrm{e}} + \lpl \sigma_{\mathrm{out}} - 
\sigma_{\mathrm{in}}  \rp
 \cdot \mathbf{n}_{\mathrm{out}} = 0,
\end{equation}
where $\boldsymbol{\rho}^{\mathrm{e}}$ is the elastic force per unit area 
exerted by the
membrane, $\sigma_{\mathrm{out}}$, $\sigma_{\mathrm{in}}$ are the stress tensor
of the flow outside and inside the capsule, and $\mathbf{n}_{\mathrm{out}}$ is 
the unit outwards
normal vector. 
Given the distribution of
$\boldsymbol{\rho}^{\mathrm{e}}$ on the membrane, the flow obeys at each time 
the Stokes equations with $N$ point forces exerted on the fluid:
\begin{eqnarray} \label{eq:stokes}
-\boldsymbol{\nabla} p + \mu \boldsymbol{\nabla}^{2}\bu & = & -\sum_{i=1}^{N} 
\bfor_{i}
\delta \lpl \bx -\bx_{i} \rp, \nonumber \\ 
\boldsymbol{\nabla} \cdot \bu \bxpp & = & 0,
\end{eqnarray}
where $p$ and $\bu$ are the pressure and velocity field.
$\bfor_{i}$ denotes the force on the fluid at the position $\bx_{i}$ and can be
approximated by
\begin{equation}
 \bfor_{i} = \int_{S_{i}}\boldsymbol{\rho}^{\mathrm{e}}_{i} d S_{i},
\end{equation}
where $S_{i}$ represents the elemental patch around the $i$th Lagrangian point. 
We solve the governing flow equations with a boundary integral method, 
accelerated by the general geometry
Ewald method (GGEM) proposed by Hern{\'a}ndez-Ortiz {\it et 
al.}~\cite{graham07_prl}. GGEM decomposes the Stokes solution into
two parts: i) the short-ranged interactions computed by traditional boundary
integral techniques; and ii)  the long-ranged interactions handled by a 
mesh-based
Stokes solver. In our implementation, we take the Stokes sub-solver of the 
open source software
NEK5000~\cite{nek5000-web-page} as the mesh-based solver. We perform
 singular and nearly-singular integration for an accurate near-field solution,  
see Ref.~\onlinecite{zhao_cell_jcp2010}.
A proper treatment of the two is crucial to achieve the necessary numerical 
accuracy.~\cite{curse_nearsing, 
zhu2013low}

The flow domain is infinite and bounded only by a plane wall.
However, in the 
framework of GGEM, a computational domain of finite size is needed;  here we 
choose $24a (x) \times 24a (y) \times 15a 
(z)$. We impose periodic boundary direction in the streamwise ($x$) and 
spanwise ($z$) direction, while keeping the
bottom wall stationary and imposing at the upper boundary a time-periodic 
velocity in the $x$ direction.
The domain is discretized by $4000$ cubic spectral elements with $4\times 4 
\times 4$ Gauss-Lobatto-Legendre 
points and the grid is refined in the region where the capsule moves. This 
strategy has been used to 
study a capsule in an unbounded shear, producing results in agreement with  
published data obtained with a traditional 
boundary integral method~\cite{lac04sphe}.

Once the flow solution is available, the velocity is known at each
Lagrangian point, $\bu \lpl  \bx_{i} \rp$.
Due to the
no-slip, non-penetrating boundary condition on the membrane of the capsule, 
the rate of change of the position of the $i$th point is given by
\begin{equation}\label{eq:update_lag}
  \frac{d \bx_{i}}{dt} = \bu \lpl \bx_{i} \rp.
\end{equation}
Given the new coordinates of the $N$ Lagrangian points, we are able to 
calculate the elastic
force per unit area on each point. The computation of the stress is based on 
the displacement
with respect to the undeformed shape of the capsule, through the constitutive 
law of the
membrane, the neo-Hookean model here. We use a global spectral method based on 
the spherical
harmonics \cite{zhao_cell_jcp2010} to represent the surface of the capsule and
to solve for the elastic force $\boldsymbol{\rho}^{\mathrm{e}}$. The advantage 
of this
approach is twofold: i) high order spatial derivatives on the material points 
are computed
with high accuracy, which is crucial for the  calculation of the elastic force;
ii) the same spectral discretization can be used for the boundary integration 
performed when
solving the short-ranged hydrodynamic interactions. For the details of our 
implementation, the
readers are referred to Refs.~\onlinecite{lai14}, \onlinecite{lailai_phd}. One 
of the important features is the capability of
simulating deformable capsules in general geometries, as illustrated by the 
design of a deformability-based cell sorting device \cite{zhu_sorting_14SM}
and a constricted micro-fluidic 
channel that can be used to infer the mechanical properties of cellular 
particles~\cite{rorai_constriction}.

\section{Results}
\subsection{Trajectories}\label{sec:traj}

\begin{figure}
        \centering
    \subfigure[]{\label{fig:tra_cap}
    \hspace{-3em}\includegraphics[scale = 0.55] 
{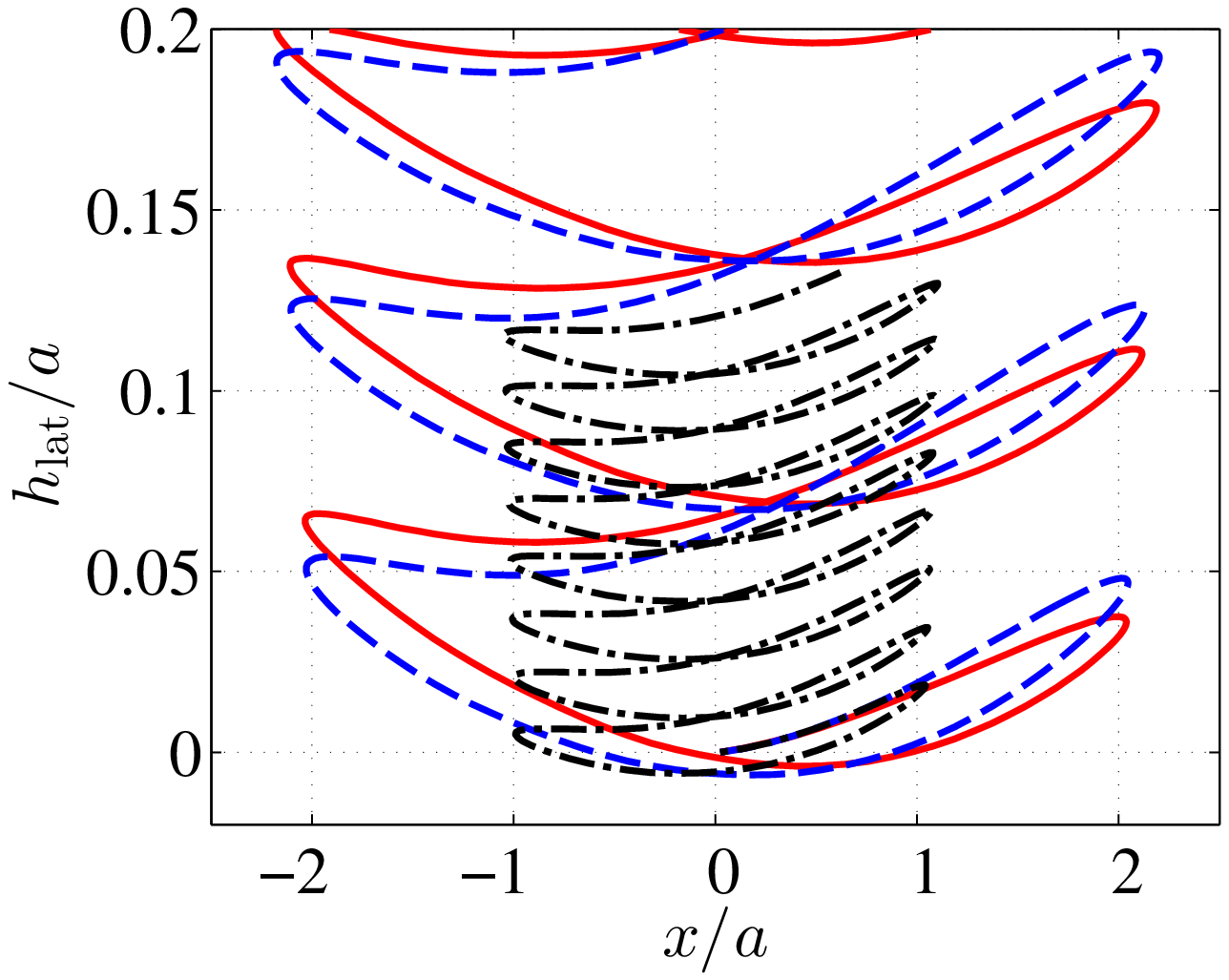}
 }
    \subfigure[]{\label{fig:tra_mig_t}
    \hspace{0em}\includegraphics[scale = 0.55] 
    {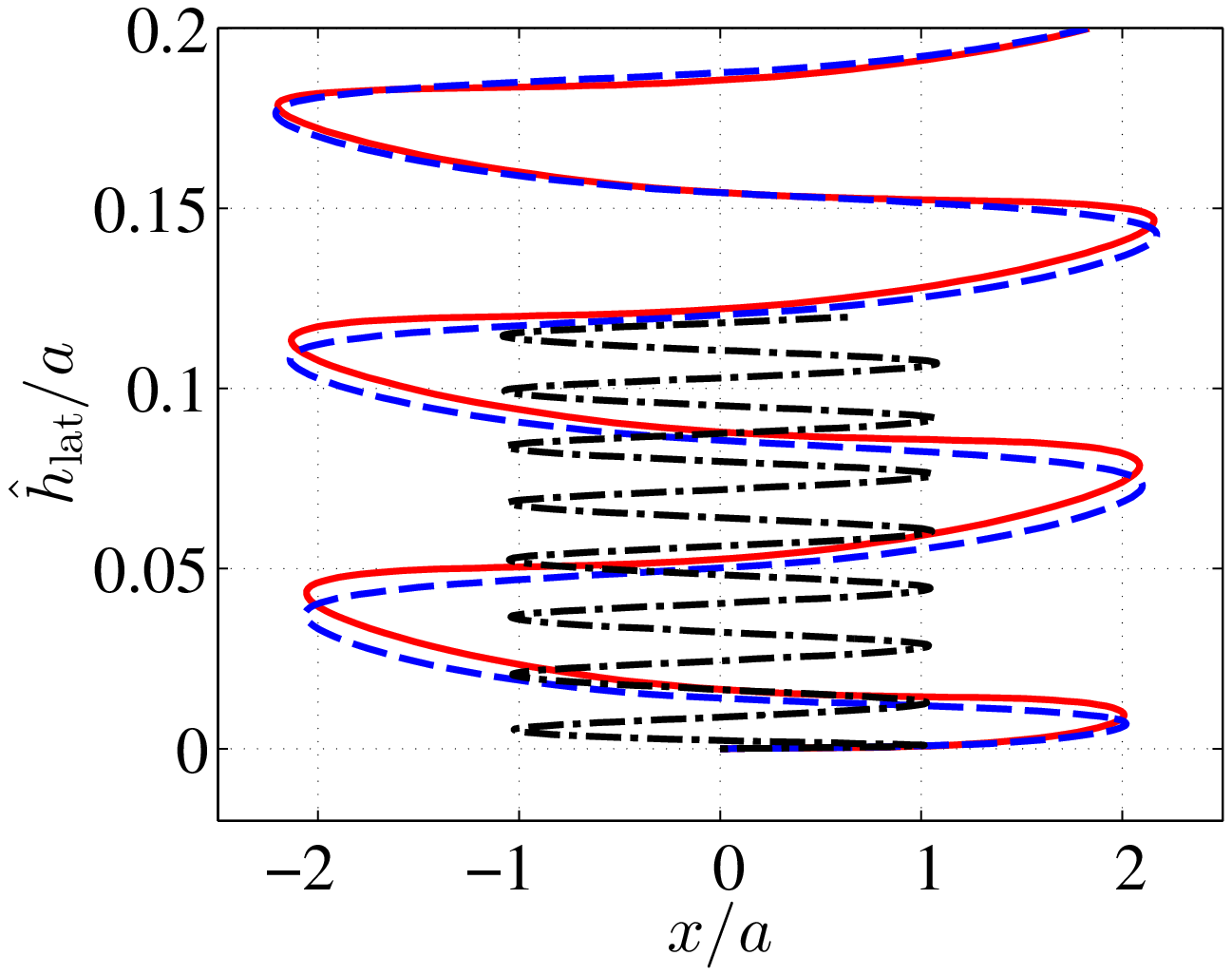}
 }
        \caption{(Color online) 
        The trajectories of  (a) $\left( x/a, \hl \right)$ of the center of 
mass, and  (b) $\left(x/a, \hat{h}_{\mathrm{lat}}/a 
\right)$ of the surface centroid 
 on the shear plane, of the capsules $\caAom=\lpl 0.15,1 \rp$ (solid), 
$\caAom=\lpl 0.3,1 \rp$ 
(dashed) and $\caAom=\lpl 0.3,2
\rp$ (dot-dashed). The capsules are released with an initial offset $\hi=2$.}
\label{PtcleMigration}
\end{figure}

\begin{figure}
        \centering
    \subfigure[]{\label{fig:tra_cap}
    \hspace{-3em}\includegraphics[scale = 0.55] 
{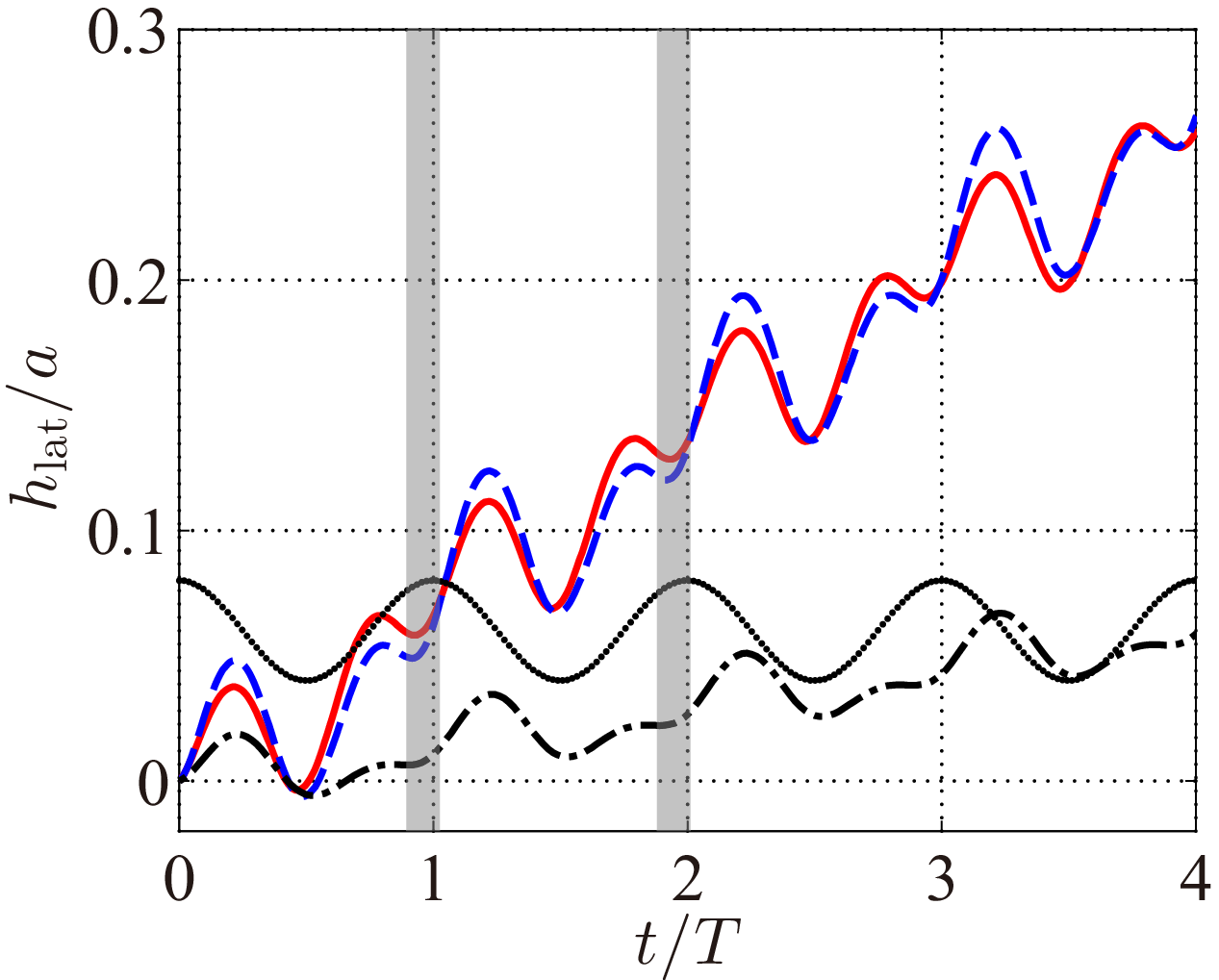}
 }
    \subfigure[]{\label{fig:tra_mig_t}
    \hspace{0em}\includegraphics[scale = 0.55] 
    {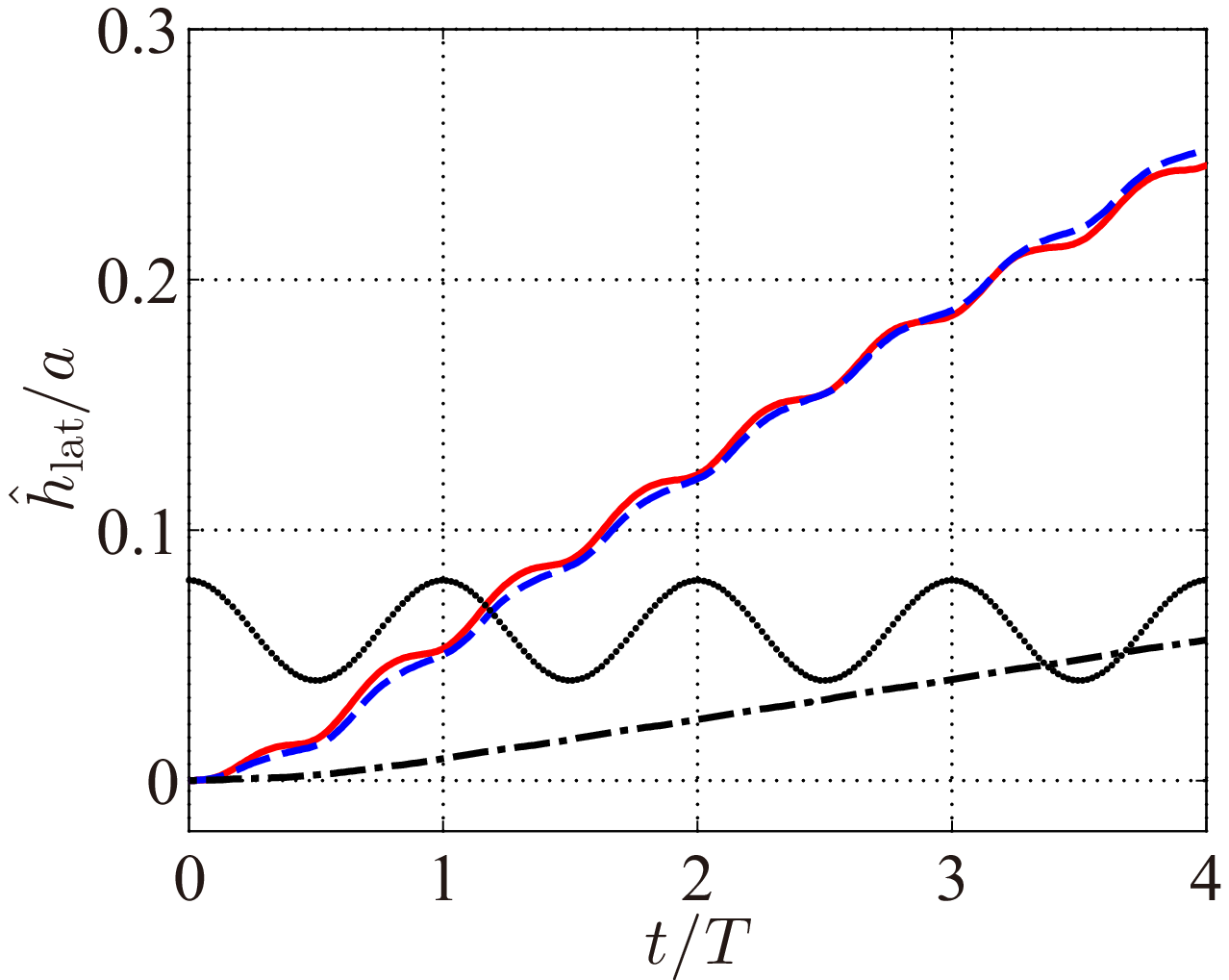}
 }
        \caption{(Color online)  (a) Lateral migration $\hl$, of the center 
of mass, and 
 (b) $\hat{h}_{\mathrm{lat}}/a$, of the surface centroid of 
        the same capsules in figure \ref{PtcleMigration}. The dotted curve 
indicates
the imposed periodic shear $\dgam=\dgamax \cos \lpl \omega t \rp$, arbitrarily 
scaled for
visualization purposes. The shaded region indicates the time when the the
instantaneous lateral migration $\hld$ and the shear $\dgam$ reach the local 
minimum and maximum respectively.}
\label{PtcleMigrationTime}
\end{figure}

We firstly look at the trajectories of the capsules. The time evolution of 
the center of mass 
and surface centroid of the capsule are depicted in the shear plane ($x-y$ 
plane)
in figure~\ref{PtcleMigration} for three cases, $\caAom=\lpl 0.15,1 \rp$,
$\caAom=\lpl 0.3,1 \rp$ and $\caAom=\lpl 0.3,2\rp$. Note that the center of 
mass is equivalent to the volume centroid,
as opposed to the surface centroid, and the two centroids coincide for 
centersymmetric objects while they
deviate as this symmetry is lost. In this study, the capsule symmetry is broken 
by the presence of the wall.
The motion is a combination of oscillations in the
streamwise ($x$) direction due to the oscillating background flow and 
wall-normal lateral
migration due to the hydrodynamic lift. The trajectories of the two centroids, 
$\hl$ and $\hat{h}_{\mathrm{lat}}$ display significant differences: the center 
of mass moves vertically in a reciprocal way while the surface centroid migrates
away from the wall continuously.

This is also illustrated in
figure~\ref{PtcleMigrationTime} where we display the time migration of the two 
centroids, $\hl$ and 
$\hat{h}_{\mathrm{lat}}/a$. They both vary periodically with the frequency of 
the imposed shear, but neither 
resembles a regular sinusoidal as the shear: the former increases with time 
non-monotonically, its migration velocity can be negative while the latter 
increases monotonically
and hence the velocity is always positive. This difference reflects the 
asymmetry of the shape of the capsule. They 
nevertheless have approximately the same time-averaged value; after all,
both centroids represent the motion of the capsule and thus coincide with each 
other in a statistical sense.
In the following, we mostly study the time-averaged capsule migration and thus 
only report results for 
the center of mass.

We further note that a solid spherical particle will not migrate  in the 
wall-normal
direction in Stokes flow, while its ellipsoidal counterpart will exhibit an 
oscillatory vertical
drift, with a zero mean. 
Nonlinearity is the cause of the particle migration, either due to inertial 
effects, a viscoelastic fluid or to  
the deformable membrane as in our case. 
As an example of solid particles, we note that a sphere has been shown to 
migrate in a constant shear flow of a 
viscoelastic 
fluid \cite{d2010viscoelasticity} due to the nonlinear relation between the 
fluid stress and the
strain rate. Under oscillating shear \cite{d2010viscoelasticity}, the spherical 
particle follows a 
wagging trajectory, very similar to that we observe here. The physical picture 
is also 
analogous to that  of a microswimmer exploiting nonlinearity to achieve a 
net locomotion despite a reciprocal motion, an 
interesting case in the low-Reynolds-number swimming dynamics \cite{lp09}. 
In the Stokes regime, but in a viscoelastic
or non-Newtonian fluid, a
reciprocal swimming pattern can lead to a net displacement \cite{pak10, 
keim2012fluid,alex14}, although the same 
strategy
would not work in a Newtonian fluid~\cite{purcell77,lp09}.

We further notice that for the same oscillation frequency, $\omgnon=1$, the 
motion of the two capsules
with $\Ca=0.15$ and $\Ca=0.3$ is similar despite the different deformability. 
Conversely, 
the trajectory of the same capsule,
$\Ca=0.3$, varies significantly when the frequency of the imposed shear 
increases from $\omgnon=1$
to $\omgnon=2$. The streamwise domain spanned by the capsule reduces to 
about half
when doubling $\omgnon$, as expected since we keep the same maximum shear. 
The capsule lateral migration also
diminishes considerably in this case, as shown in figure~\ref{fig:tra_mig_t}.

\subsection{Velocity of the capsule}

\begin{figure}
        \centering
    \subfigure[]{\label{fig:ulat_t}
    \hspace{-3em}\includegraphics[scale = 0.55] {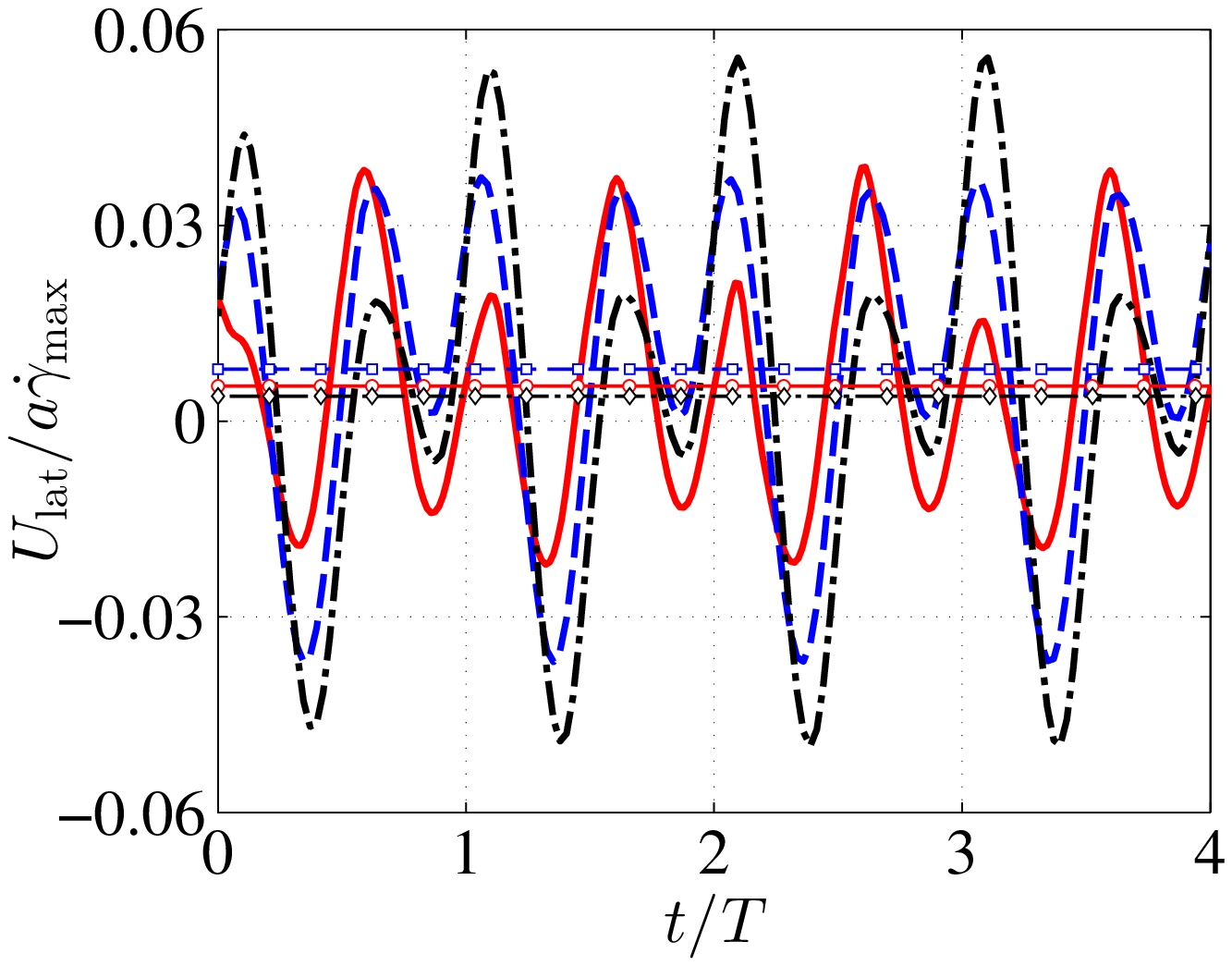}\hspace{-1em}
 }
    \subfigure[]{\label{fig:uslip_t}
    \includegraphics[scale = 0.55] {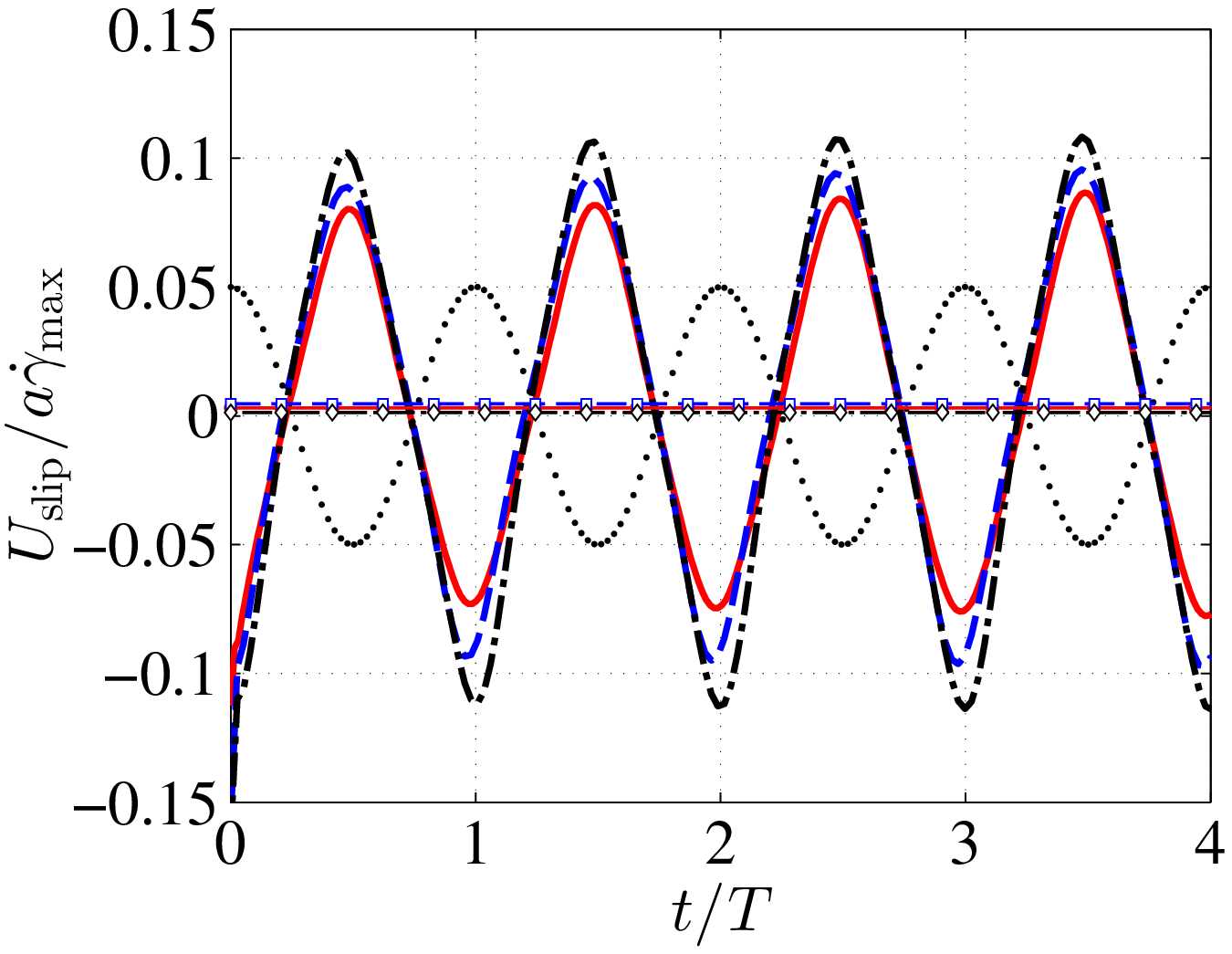}
 }       
        \caption{(Color online) \subref{fig:ulat_t}: Lateral migration velocity 
$\ul$  and \subref{fig:uslip_t}: slip 
velocity $\us$,  as a function of the dimensional time $t/T$ for capsules with 
$\Ca=0.0375$ (solid curve), $\Ca=0.15$ (dashed) and
$\Ca=0.6$ (dot-dashed), released with an initial offset $\hi=2$; the 
oscillation frequency of the imposed shear 
$\omgnon 
= 
5/3$. The time-averaged migration and slip
velocities are denoted by horizontal lines with circles ($\Ca=0.0375$), squares
($\Ca=0.15$) and diamonds ($\Ca=0.6$). The dotted curve indicates
the applied oscillating shear $\dgamax \cos \lpl \omega t \rp$, arbitrarily 
scaled for
visualization purposes.}
\label{fig:ulat}
\end{figure}

In this section, we analyze the migration and slip velocity of the capsule. The 
lateral migration velocity $\uld$, 
is computed as $\uld=d \hcw /dt$ and the slip velocity $\usd = \dgam \hcw - 
U_{x}$ where $\hcw=\hid+\hld$ is the 
distance between the capsule and the wall and $U_{x}$ the streamwise velocity 
of the center of the 
capsule. The 
temporal evolution of $\ul$ and $\us$ are depicted in figure~\ref{fig:ulat_t} 
and figure~\ref{fig:uslip_t}
for three capsules with the same initial offset $\hi=2$ and frequency of the 
imposed shear, $\omgnon=5/3$. After 
an 
initial
transient, the two velocities vary periodically in time. The transient is also 
observed in the case of steady shear~\cite{singh2014lateral}, before the 
capsule 
reaches a quasi-steady state; only then its deformation and velocity do not 
vary with the 
instantaneous capsule-wall distance $\hcw$. In the case of periodic shear 
investigated here, 
the capsule motion reaches a quasi-periodic regime. Unless otherwise  
specified, we only examine its dynamics at 
this stage, when its net migration $\hld$ is 
negligible compared to the initial offset $\hid$ and $\hid \approx \hcw$.  In 
the following, 
$\hid$ will therefore be used to denote the 
capsule-wall distance.

As clear from figure~\ref{fig:uslip_t}, the slip velocity and the background 
shear flow have 
different signs. This indicates that the capsule moves faster than the local 
background flow, an effect more pronounced 
when $t/T$ approximately assumes integer values and the capsule returns about 
the initial location ($x=0$).
This is different to what is observed in a steady shear flow where the capsule 
always lags behind the
surrounding fluid~\cite{singh2014lateral,nix2014lateral}. In fact, as $t/T 
\approx 1, 2, 3...$ (shaded 
regions in figure~\ref{fig:tra_mig_t}),  
 the capsule-wall distance $\hcw = \hid + \hld$ is at a local minimum and the 
background 
shear $\dgam$ reaches its maximum value. At this time, the background velocity 
$\dgam \hcw$, the product of the two, 
decreases because the phase difference is around $\pi$. This phase 
difference does not exist 
in steady shear when there are no delays and the local background 
flow depends only on the capsule-wall distance $\hcw$.

\begin{figure}
        \centering
    \includegraphics[width=0.5\columnwidth] {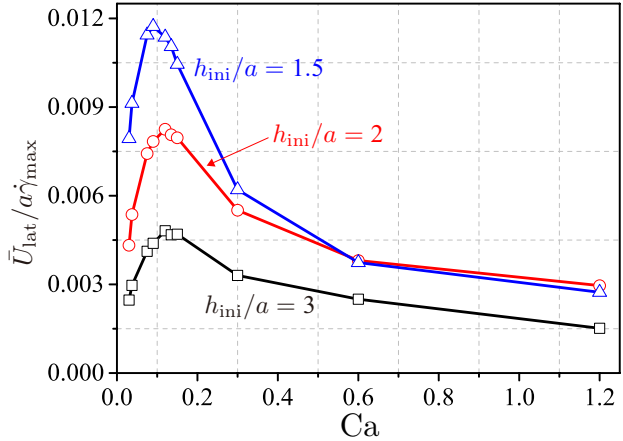}\hspace{0em}       
    \caption{(Color online) Lateral migration velocity $\ul$ as a function of 
the
dimensional time $t/T$ and the capillary number $\Ca$, at fixed oscillation 
frequency $\omgnon = 5/3$. 
Data are reported for three initial offsets, $\hi=1.5$, $\hi=2$ and $\hi=3$ 
indicated by
the triangles, circles and squares respectively.}
\label{fig:ulat_ca}
\end{figure}

The time-averaged value of the migration velocity $\uml$ and of the slip 
velocity 
$\us$ of the three capsules are indicated in figure~\ref{fig:ulat} 
by horizontal lines. For all cases, the mean migration velocity 
is positive:
the capsule undergoes a net migration away from the wall. However, the mean 
slip velocity is almost zero, 
namely, the capsule has no net motion in the streamwise direction. The capsule 
with the intermediate 
capillary 
number $\Ca=0.15$ has a higher migration velocity 
 than its floppy ($\Ca=0.0375$) and stiff ($\Ca=0.6$) 
counterparts. 
To further examine this dependence, the migration velocity $\uml$ is therefore 
displayed versus
the capillary number $\Ca$ in figure~\ref{fig:ulat_ca}  for three different 
initial offsets. 
The mean velocity varies
non-monotonically with $\Ca$, the optimal capillary number $\Caopt$ being 
around $0.1$. This non-monotonic 
dependence of the migration velocity is in contrast to what is observed in 
steady shear, 
where the 
lift velocity increases monotonically with $\Ca$ 
\cite{ma2005theory,pranay2012depletion,singh2014lateral}.

In steady flows, the particle lateral migration velocity is a linear function of
the normal stress differences $N_{1}-N_{2}$ as shown among others in 
Ref.~\onlinecite{ma2005theory} and \onlinecite{pranay2012depletion}.
It is thus natural to investigate whether this relation holds in an oscillating 
shear as well. The normal stresses 
$N_{1}$ and $N_{2}$ are defined 
as,
\begin{align}\label{eq:normsts}
 N_{1} = \sigma_{xx} - \sigma_{yy},\\ \nonumber
 N_{2} = \sigma_{yy} - \sigma_{zz},
\end{align}
where $\bsg$ denotes the extra stress due to the presence of the capsule in the 
volume $V$:
\begin{equation}
 \sigma_{ij} = -\frac{1}{V}\int_{S}\boldsymbol{\rho}^{\mathrm{e}}_{i}x_{j}dS.
\end{equation} 
We consider the nondimensional stress difference $\lpl N_{1}-N_{2} \rp/\mu 
\phiv \dgamax$, with
$\phiv=1/V$ the volume fraction; in the dilute limit, the normal stress 
difference $\DN = \lpl N_{1}-N_{2}\rp / \mu
\phiv \dgamax$ 
represents the capsule-induced stress normalized by the viscous stress.

\begin{figure}
        \centering
            \subfigure[]{\label{fig:nmsts_ulat_0075}
    \hspace{-3.5em}\includegraphics[scale = 0.36] 
{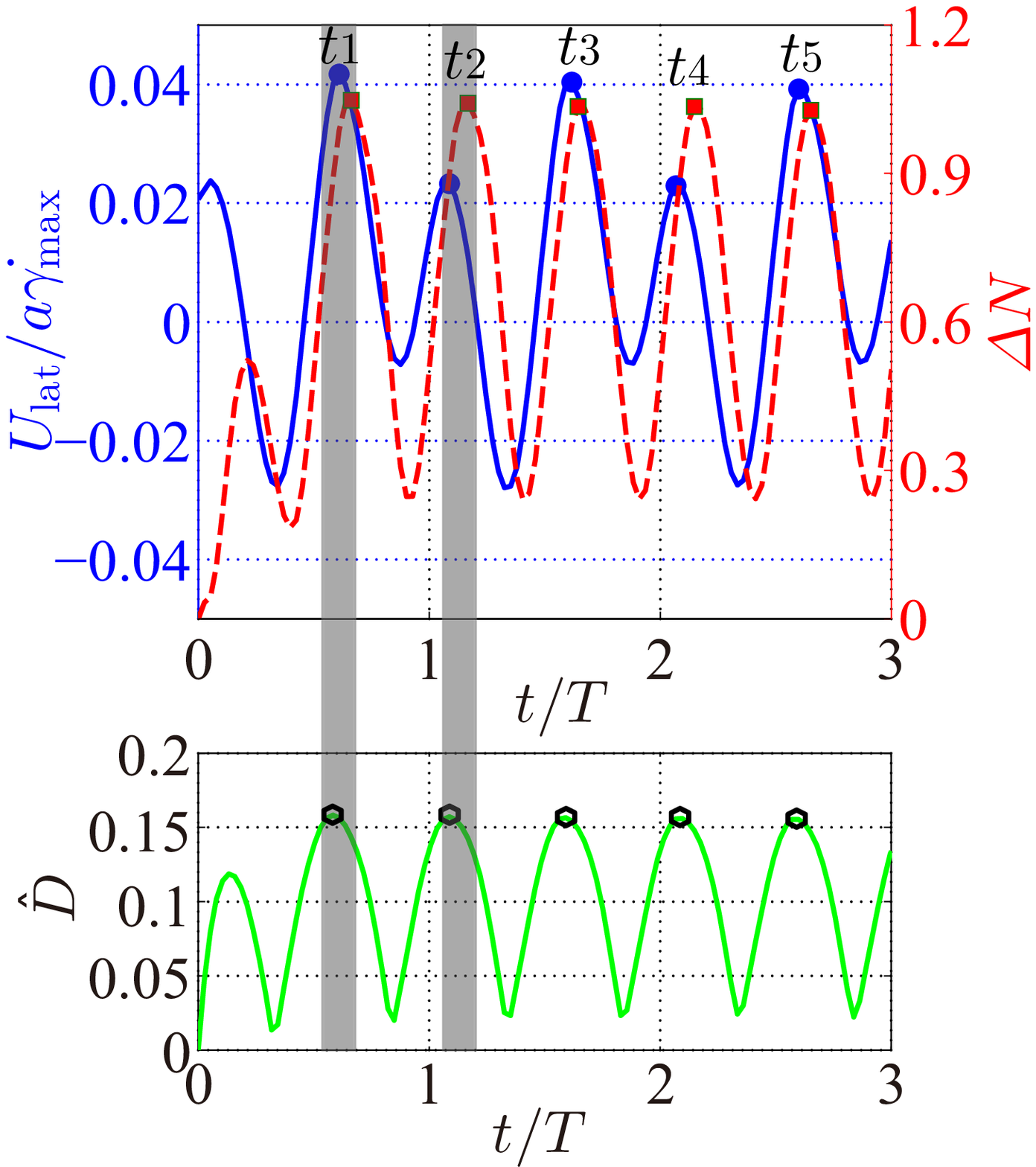}\hspace{-0.75em}
 }  
    \subfigure[]{\label{fig:nmsts_ulat_012}
    \includegraphics[scale = 0.36] 
{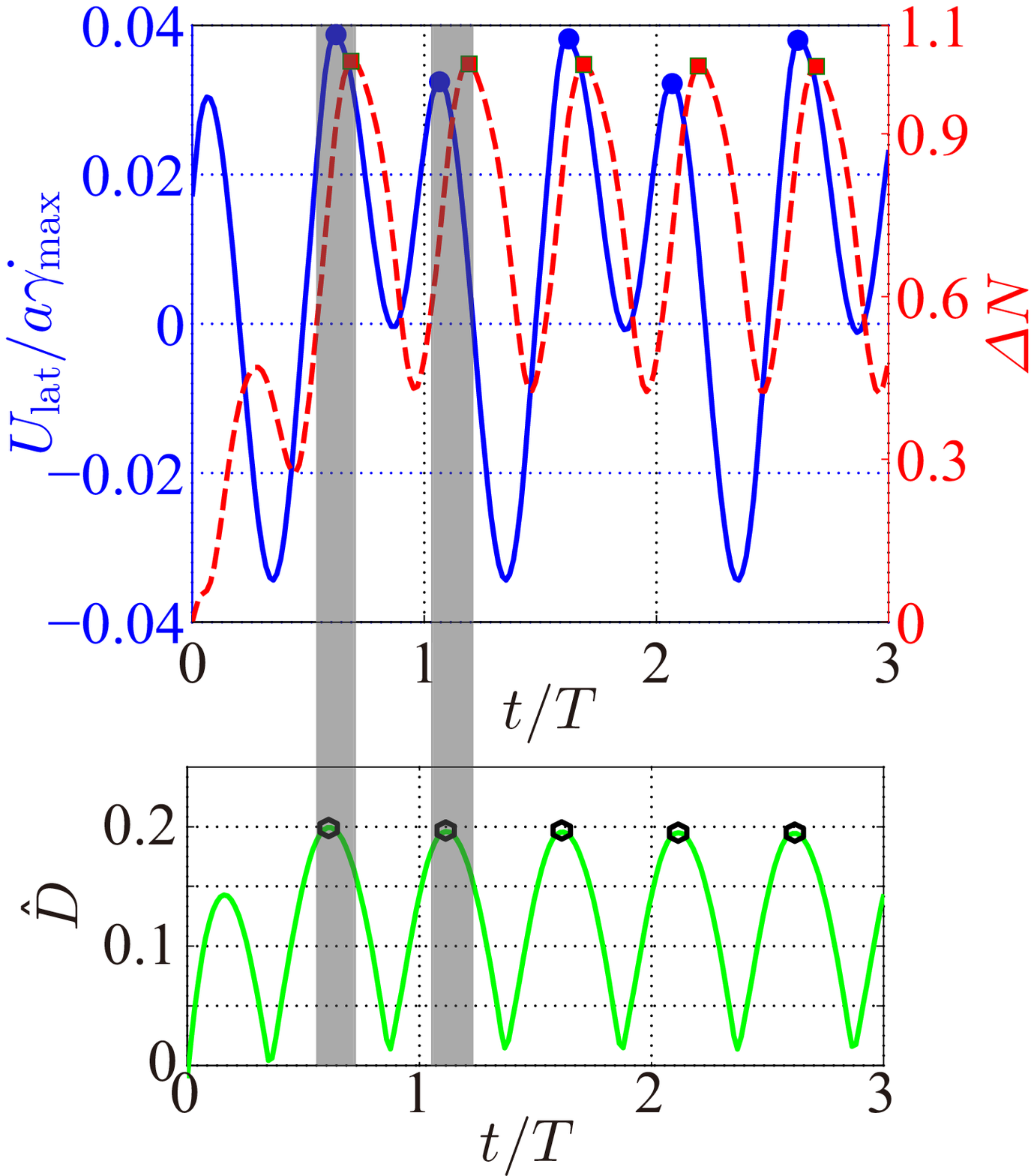}\hspace{-0.75em}
 }     \subfigure[]{\label{fig:nmsts_ulat_06}
    \includegraphics[scale = 0.36] {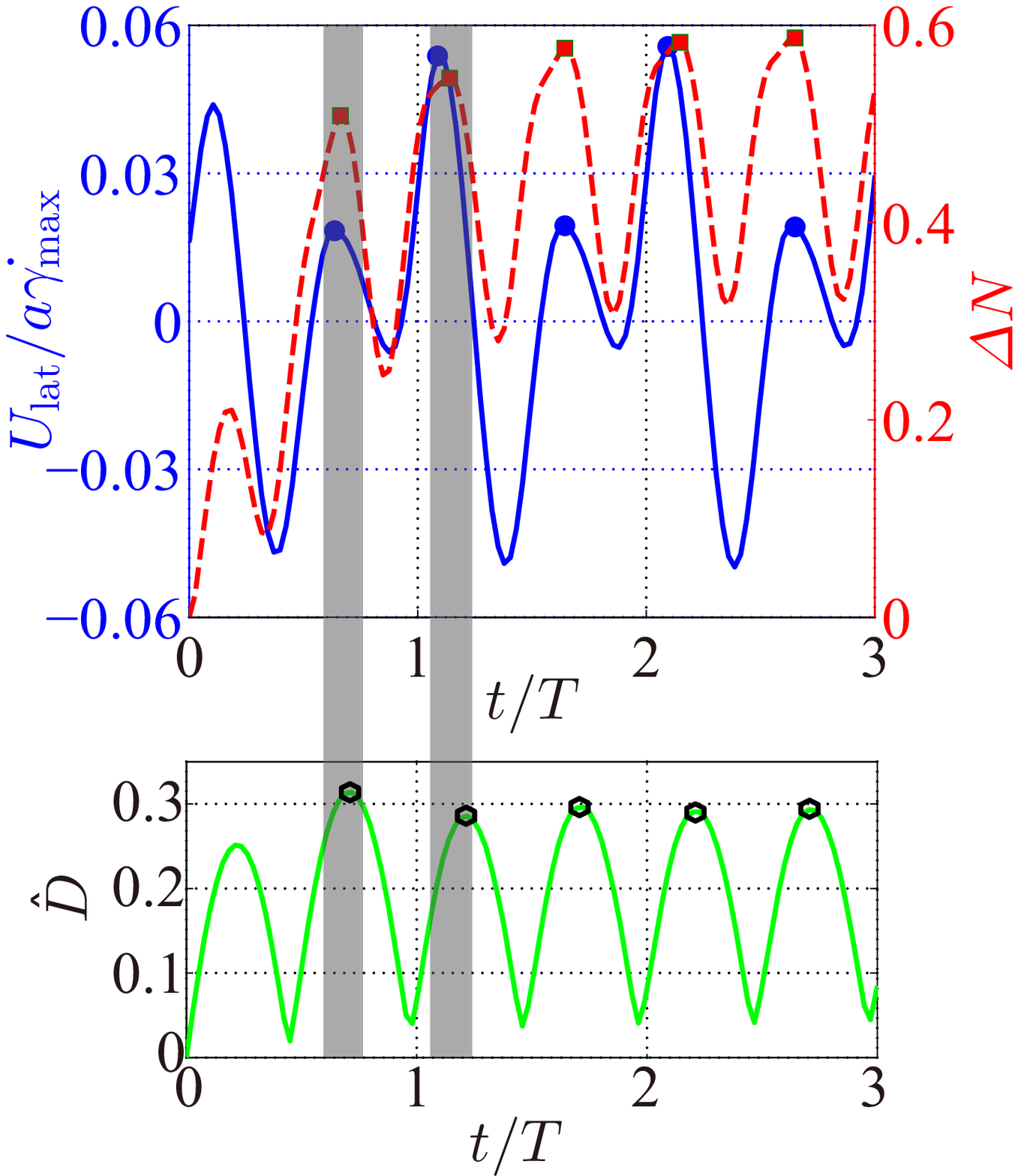}
 }
        \caption{(Color online) Top panel: lateral migration velocity, $\ul$, 
(solid lines) and
normal stress difference $\Delta N$ (dashed lines) versus the nondimensional 
time $t/T$, for three capillary numbers
$\Ca=0.075$ \subref{fig:nmsts_ulat_0075}, $0.12$ \subref{fig:nmsts_ulat_012} 
and 
$0.6$ \subref{fig:nmsts_ulat_06}. 
Bottom panel: deformation index $\hD$ versus $t/T$ for the same capsule.
The initial offset is $\hi=2$ and the frequency of the imposed shear
$\omgnon
= 5/3$. 
The local maxima of the velocity, stress and deformation are marked by the 
circles,
squares and hexagons respectively; the shaded region indicates the period when 
they co-appear.}
\label{fig:nmsts_ulat}
\end{figure}

The temporal evolution of the migration velocity $\ul$ and the normal stress 
difference $\DN$ are depicted 
in the top panel of figure~\ref{fig:nmsts_ulat} for capsules with $\Ca=0.075$, 
$\Ca=0.12$ and $\Ca=0.6$. After the 
transient period, both quantities vary periodically and reach their local 
maxima with small phase
delays, which indicates a considerable correlation between these two 
quantities. The period is around $T/2$, indicating
that $\uld$ and $\DN$ change twice as faster as the oscillating shear.

We next examine the local maxima of $\ul$ (marked by circles in the figure) and 
of $\Delta N$ (indicated by squares), 
always neglecting the
peaks in the transient period. For the stiff capsule $\Ca=0.075$ (cf.\ 
figure~\ref{fig:nmsts_ulat_0075}), we identify 
five instants when
$\ul$ and $\Delta N$ both attain local maxima. At times $t_{1}$, $t_{3}$ and 
$t_{5}$,
the peaks of the two quantities have a similar magnitude with respect to their 
own scales, while at the time
$t_{2}$ and $t_{4}$, their relative magnitude is significantly different.
In other words, the ratio between the magnitude of the two  local maxima varies 
with time, something which becomes more 
evident
for the floppy capsule $\Ca=0.6$. 
In contrast, 
the two quantities $\ul$ and $\Delta N$ have
roughly the same magnitude every time they reach a local maxima for the capsule 
with $\Ca=0.12$, the one displaying the 
largest lateral migration velocity (see
figure~\ref{fig:nmsts_ulat_012}): the ratio of the peak values remains
almost constant in time. This implies that the correlation between the two 
quantities is stronger when $\Ca \approx 0.1$ 
and decreases for larger and smaller values of the capillary number 
($\Ca=0.075$ and $0.6$ in the figure). 
This high correlation suggests that the normal stress difference 
 contributes to a more pronounced lateral migration.

\begin{figure}
        \centering
             \subfigure[]{\label{fig:meansts_meanvel_ca_hi1.5}
    \hspace{-3em}\includegraphics[scale = 0.37] {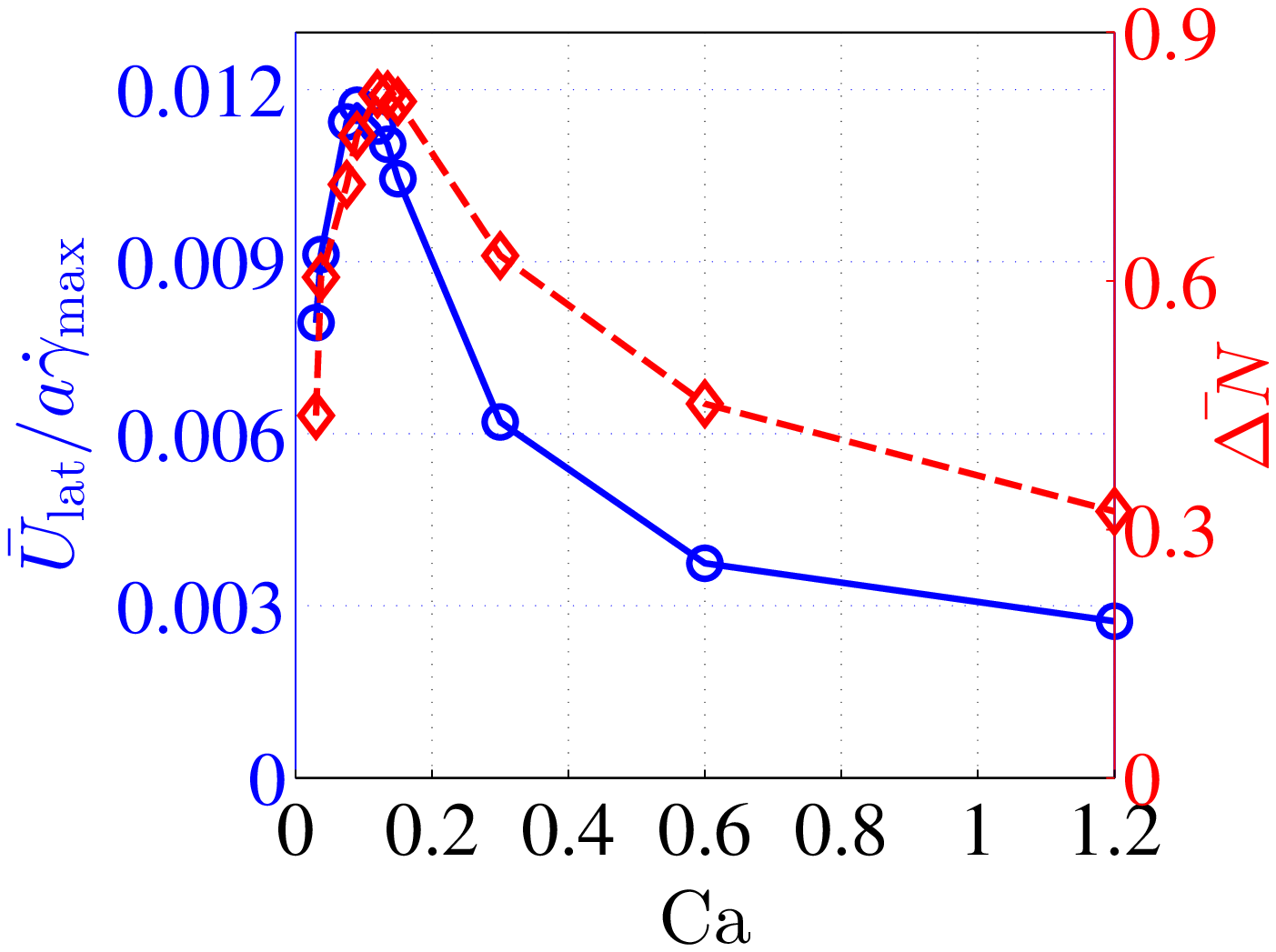}\hspace{0em}
 }
    \subfigure[]{\label{fig:meansts_meanvel_ca_hi2}
    \includegraphics[scale = 0.37] {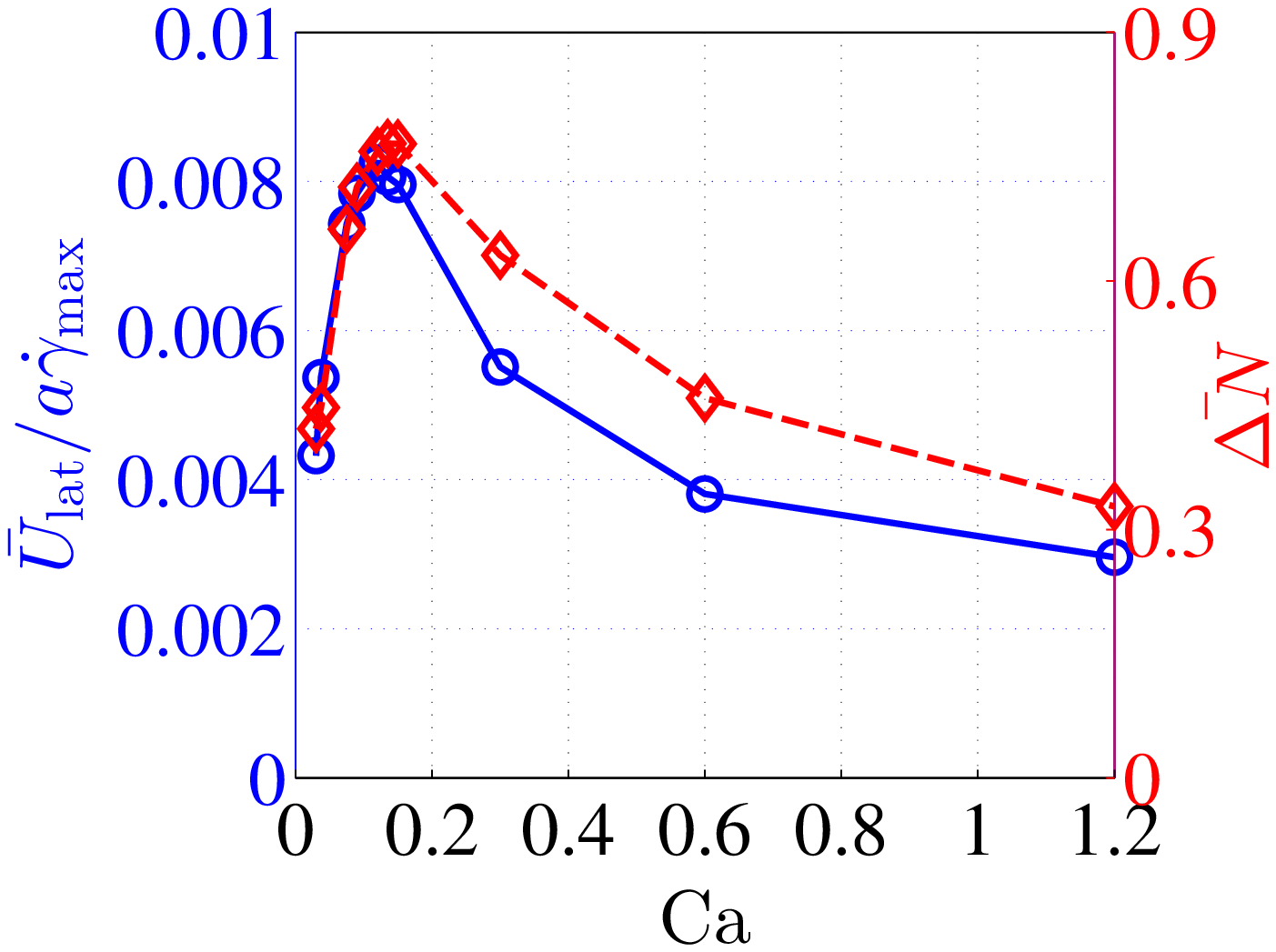}\hspace{0em}
 }     \subfigure[]{\label{fig:meansts_meanvel_ca_hi3}
    \includegraphics[scale = 0.37] {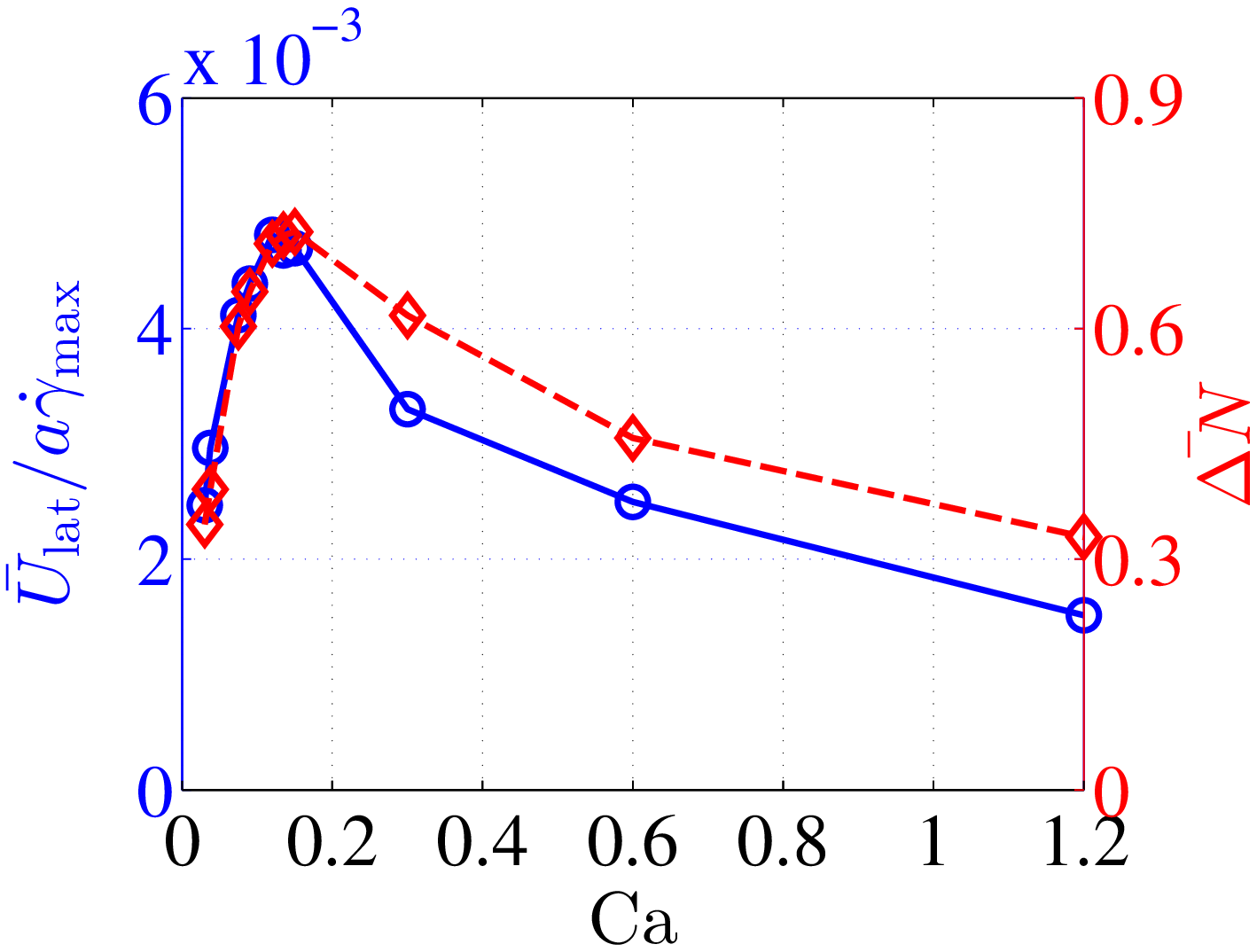}
 }     
        \caption{(Color online) Time-averaged value of the lateral migration 
velocity $\uml$ and of the normal stress 
differences
$\bar{\Delta {N}}$, versus the capillary number $\Ca$. $\uml$ and $\bar{\Delta 
{N}}$ are indicated by
the solid (resp.~dashed) curve with circles (resp.~diamonds),
measured by the left
(resp.~right) scale. The frequency of the shear is $\omgnon=5/3$ and the 
results 
are shown for  ~\subref{fig:meansts_meanvel_ca_hi1.5} $\hi=1.5$,
~\subref{fig:meansts_meanvel_ca_hi2} $\hi=2$ and 
~\subref{fig:meansts_meanvel_ca_hi3} $\hi=3$.}
\label{fig:meansts_meanvel_ca}
\end{figure}

To gain further insight,
we show the time-averaged migration velocity $\uml$ together with the 
time-averaged  normal
stress difference $\nstsm$ versus the capillary number $\Ca$ in 
figure~\ref{fig:meansts_meanvel_ca}; the data reported 
pertain to three different values of the initial distance from the wall.
Interestingly, these two quantities follow closely the same trend: a sharp 
increase 
with $\Ca$ for stiff capsules, a maximum at $\Ca \approx 0.1$ and then slower 
decrease for softer and softer membranes.

\begin{figure}
        \centering
    \includegraphics[width=0.5\columnwidth] {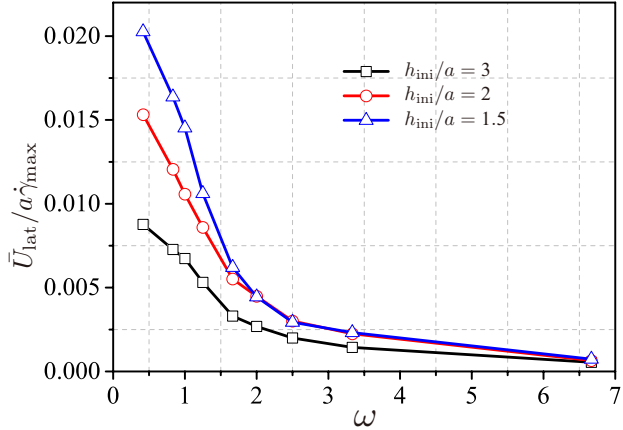}\hspace{0em}     
    \caption{(Color online) Time-averaged migration velocity $\uml$, as a 
function of the
frequency of the oscillating shear $\omgnon$. The capillary number $\Ca$ of 
the capsule is
$0.3$ and three initial positions of the capsules are chosen, $\hi=3$ denoted 
by squares,
$\hi=2$ by circles and $\hi=1.5$ by triangles.}
\label{fig:ulat_mean_omega}
\end{figure}

The parameter study is continued by examining the dependence of the lateral 
migration velocity on the frequency of 
oscillation of the imposed shear, $\omgnon$. 
Figure~\ref{fig:ulat_mean_omega} displays the time-averaged
migration velocity, $\uml$, versus $\omgnon$ for the capsule with $\Ca=0.3$ 
and the three offsets 
 $\hi=1.5$, $2$ and $3$. 
As $\omgnon< 2$, $\uml$ decreases sharply with $\omgnon$, almost 
linearly;
for $\omgnon >2$, the migration velocity decreases more slowly, approaching
asymptotically zero for large $\omgnon$. 
The cases with $\omgnon=0$ correspond to a capsule in steady
shear flow. Conversely, as $\omgnon \rightarrow \infty$, the flow 
oscillations are much faster than the
relaxation time of the capsule and the capsule therefore does not have 
sufficient time to adjust to the flow
before it changes direction: the deformation and the consequent lateral 
migration are hence negligible. 

\begin{figure}
        \centering
    \includegraphics[width=0.5\columnwidth] 
{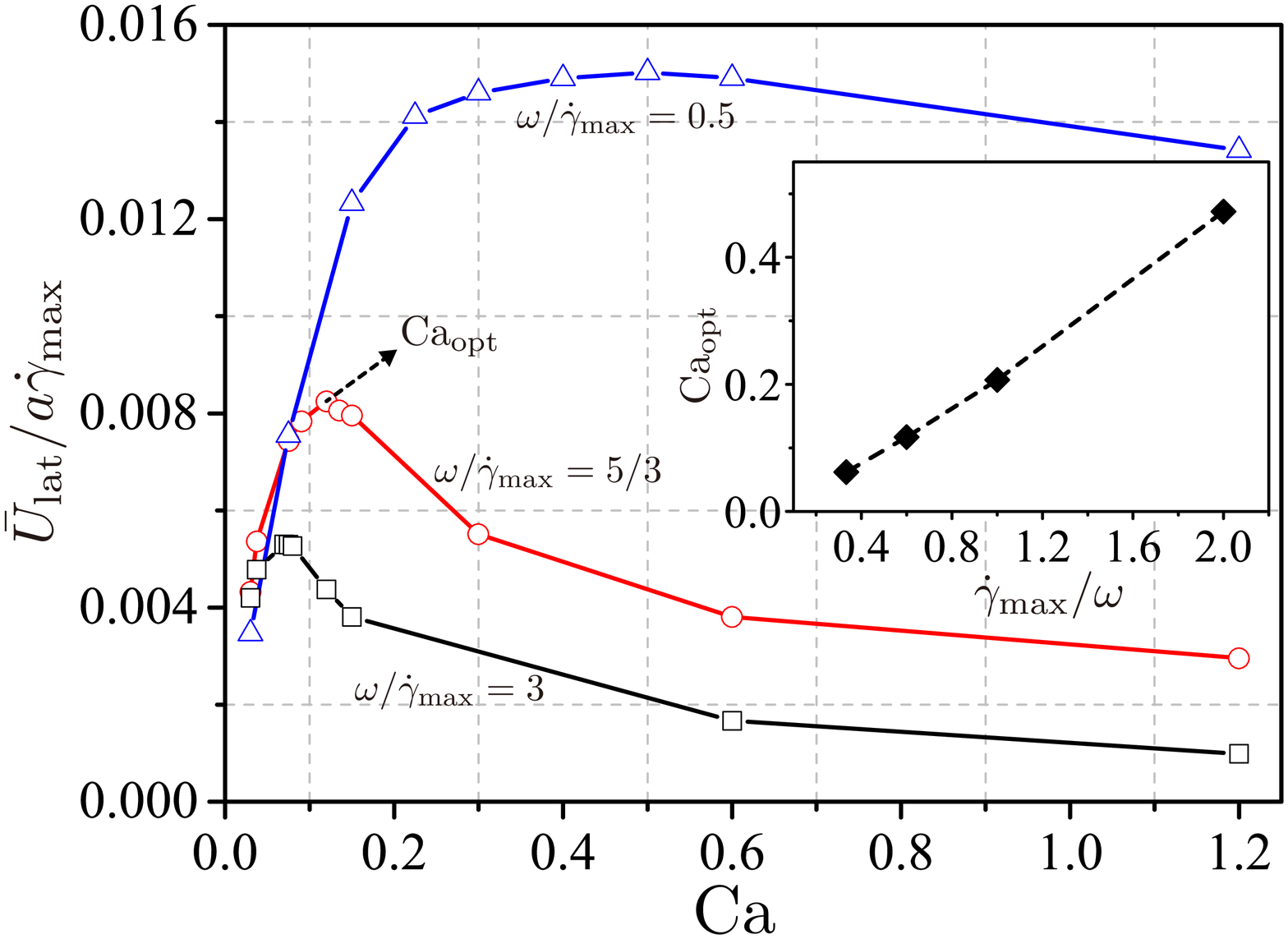}\hspace{0em}       
    \caption{(Color online) The time-averaged lateral migration velocity $\uml$ 
versus the capillary number $\Ca$, for 
three shear 
frequencies, $\omgnon=0.5$ (triangles), $\omgnon=5/3$ (circles) and 
$\omgnon=3$ (squares), the 
initial offset $\hi=2$. $\Caopt$ indicates the capillary number of the capsule 
with the maximum migration 
velocity 
 and the inset shows its dependence on 
$\dot{\gamma}_{\mathrm{max}}/\omega$.}
\label{fig:umean_ca_threeomega}
\end{figure}

The variations of the time-averaged
migration velocity $\uml$ with the oscillation frequency are displayed in 
figure~\ref{fig:umean_ca_threeomega}  versus 
the capillary number $\Ca$ for capsules with
an initial offset $\hi=2$. The non-monotonic variation of 
$\uml$ with $\Ca$ holds for all cases, but the capillary number $\Caopt$ of
maximum lateral velocity changes with the frequency $\omgnon$. 
The inset of the same figure shows that $\Caopt$ varies almost linearly with 
the inverse of the shear frequency $\dot{\gamma}_{\mathrm{max}}/\omega$. In 
fact, as the migration velocity, 
normal stress difference and
capsule deformation vary at a frequency of roughly $2\omgnon$ (see 
figure.~\ref{fig:nmsts_ulat}), 
$2\omgnon$ can be considered as the effective frequency of the flow 
oscillation 
determining the capsule behavior.
Accordingly, the ratio of the relaxation time of the capsule over the effective 
time scale of the 
flow is $2\Ca\omgnon$. 

The inset indeed shows that 
$2\Caopt\omgnon \approx 0.5$.
In another words, lateral migration becomes strongest 
as  the effective time scale of the flow is of the order of the relaxation time 
of the elastic membrane.

It is worth noting that this scaling has similarities with that observed in 
many other situations
 where the flow dynamics is significantly influenced by the nonlinearity solely 
arising from the elasticity of 
 the fluid or structure. A rotating helical slender body, a typical model 
micro-swimmer propelling 
like a cork-screw,  
attains 
the most efficient propulsion when the relaxation 
time of the polymeric fluid it is  immersed in is of the order of the typical 
flow 
time~\cite{LiuHelixPnas,spagnolie2013locomotion}. 
Similarly, an elastic filament~\cite{wiggins1998flexive,tony2006experimental} 
or 
flapper~\cite{arco2014viscous} actuated in a viscous fluid for propulsion or 
pumping reaches maximum  
efficiency as the so-called sperm number~\cite{machin1958wave}, the ratio of 
the time scale of the elastic structure 
over 
that of the flow, is of order one. We further note that the relation between 
the optimal capillary number
and the frequency of the applied shear could be potentially used to aid the 
design of flow-assisted devices
to sort deformable 
cells~\cite{shapiro2005practical,ye2014effects,krueger2014deformability}. 
Specific cells may 
be extracted from a dilute suspension in an oscillating Couette device if the 
frequency of the applied shear is tuned 
to match  the capillary 
number of the targeted cells; these cells would in fact migrate with highest 
velocity. The frequency of the applied shear would have to depend on the 
deformability of the targeted cells, 
therefore the same device can be used to sort cells with different 
deformabilities by varying the frequency of 
operation.

Furthermore, the relation $2\Caopt\omgnon \approx 0.5$ can also be 
implied theoretically in the limit of high-frequency 
oscillation, i.e. $\omega/\dot{\gamma}_{\mathrm{max}} \gg 1$, when the capsule 
undergoes small deformation. We can assume the capsule reaches the peak 
deformation and migration velocity at the same capillary number
$\Caopt$. In an unbounded oscillating shear, the leading order of peak 
deformation is~\cite{matsunaga2015deformation} $\hDpk = 
\frac{1}{2\omega/\dot{\gamma}_{\mathrm{max}}}$ and that of the equilibrium 
value $D$ in the steady case is~\cite{dbb1980_asym, li2008front} $D = 
\frac{25}{12}\Ca$. Hence the capsule reaches the peak deformation and migration 
as  $2\Caopt \omega/\dot{\gamma}_{\mathrm{max}} = \frac{12}{25} \approx 0.5$.

\begin{figure}
        \centering
    \subfigure[]{\label{fig:nmsts_off10_omg05}
    \hspace{-3em}\includegraphics[scale = 0.435] 
{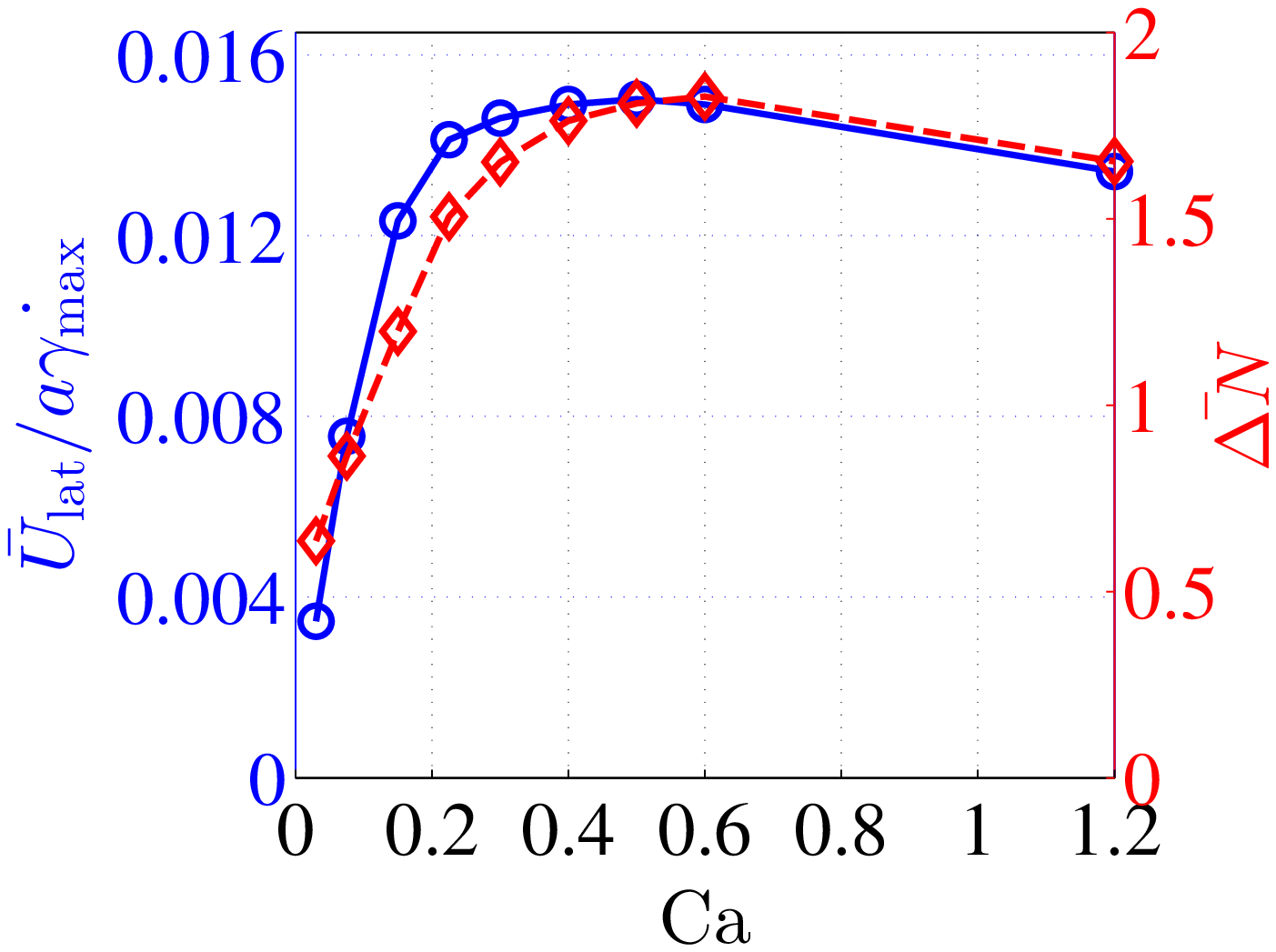}
 }
    \subfigure[]{\label{fig:nmsts_off10_omg3}
    \hspace{0em}\includegraphics[scale = 0.45] 
    {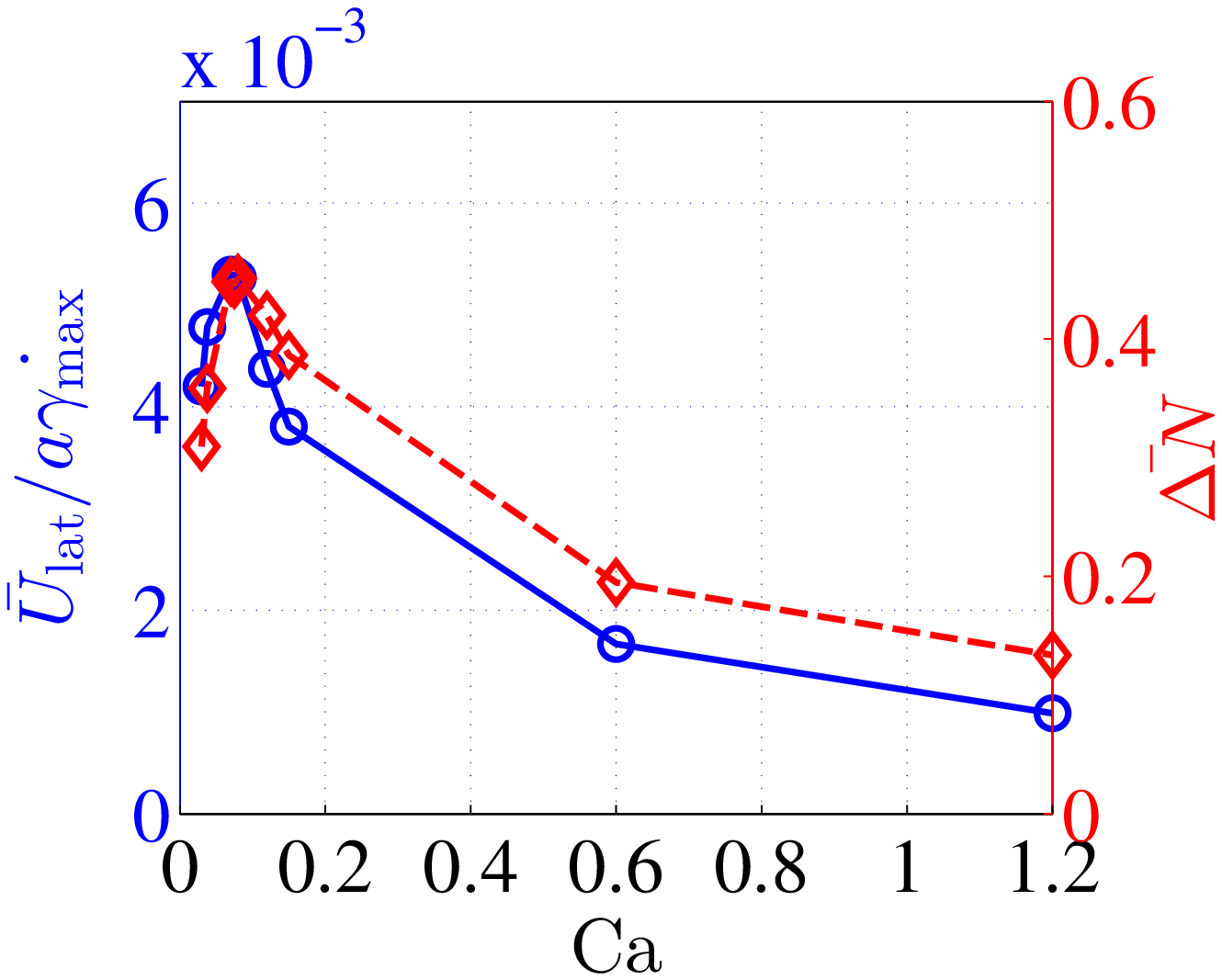}
 }
        \caption{(Color online) The time-averaged value of the migration 
velocity $\uml$ and that of the normal stress 
differences $\bar{\Delta {N}}$ versus the capillary number $\Ca$. The initial
offset is $\hi=2$:  ~\subref{fig:nmsts_off10_omg05}: $\omgnon=0.5$ and 
~\subref{fig:nmsts_off10_omg3}: $\omgnon=3$.}
\label{nmsts_vel_ca_varyomg}
\end{figure}

Finally, we report in figure~\ref{nmsts_vel_ca_varyomg} the time-averaged 
migration velocity
$\uml$ and the normal stress difference versus the capillary number $\Ca$ for 
two different oscillation frequencies,
$\omgnon=0.5$ and $\omgnon=3$. As shown in 
figure~\ref{fig:meansts_meanvel_ca}, we can clearly identify a 
positive 
correlation between these two quantities.

\subsection{Deformation of the capsule}

\begin{figure}
        \centering
    \includegraphics[width=0.6\columnwidth]
{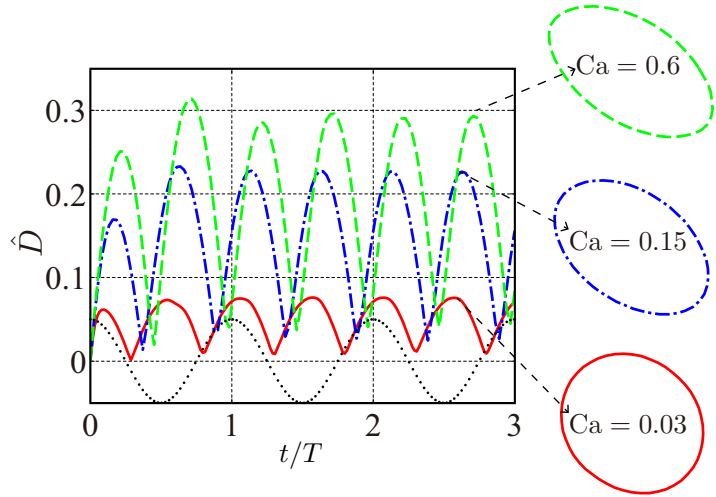}\hspace{0em}      
    \caption{(Color online) Temporal evolution of the deformation index $\hD$ 
for
capsules with initial offset $\hi=2$ and frequency of the shear 
$\omgnon=5/3$. 
Results for $\Ca=0.03$, $\Ca=0.15$ and
$\Ca=0.6$ are indicated by the solid, dot-dashed and dashed lines respectively. 
The dotted
curve denotes the oscillating background shear $\dgamax \cos \lpl \omega t 
\rp$, arbitrarily scaled for a better
visualization. The profiles of the capsules reaching the peak deformation are 
displayed on the shear plane. }
\label{fig:dindex_v_time_omega1.67}
\end{figure}

A capsule typically evolves into a prolate when subject to an unbounded shear 
flow. Hence, its
deformation is usually quantified by the so-called Taylor parameter 
$D = (L_{\mathrm{max}} - L_{\mathrm{min}})/(L_{\mathrm{max}} +
L_{\mathrm{min}})$~\cite{bar+wal+sal10},
where  $L_{\mathrm{max}}$ and $L_{\mathrm{min}}$ are the length of the major 
and minor axis 
of the elliptical profile in the shear plane. 
In our case, the profile of the capsule on the shear plane is not a ellipse, 
since the symmetry
is broken by the presence of the wall as also observed by Nix {\it et 
al.}~\cite{nix2014lateral}. 
We thus introduce the deformation index, $\hD = (L^{\prime}_{\mathrm{max}} -
L^{\prime}_{\mathrm{min}})/(L^{\prime}_{\mathrm{max}} + 
L^{\prime}_{\mathrm{min}})$, where
$L^{\prime}_{\mathrm{max}}$ is the maximum distance measured from the capsule 
surface to the its
center of mass and $L^{\prime}_{\mathrm{min}}$ is the minimum distance.
The deformation index $\hD$ quantifies how far away the capsule is from its 
stress-free shape, a
sphere in the current case.

The time evolution of the deformation index $\hD$  is illustrated in
figure~\ref{fig:dindex_v_time_omega1.67} for three values of the capillary 
number, $\Ca=0.03$, $0.15$ and $0.6$,  
same initial offset, $\hi=2$, and shear frequency $\omgnon=5/3$. 
The index $\hD$ displays wake-like variations in time. An initial overshoot 
is observed for the soft capsules, $\Ca=0.3$ and $0.6$, similar to that 
observed in the wall-bounded 
steady shear flow~\cite{singh2014lateral}. After one or two 
periods, the amplitude of the oscillations becomes constant. 
The initial transient  is more evident for the 
relatively soft capsules, $\Ca=0.15$ and
$0.6$; indeed, a higher capillary number implies a longer relaxation time, 
hence the capsule
needs more time to adapt to the unsteady flow. 
We also note in the figure that the frequency of the periodically varying 
deformation index $\hD$ is twice
that of the background shear. The stiffest capsule, $\Ca=0.03$, displays the 
maximum deformation roughly after the
local extrema of the shear and the minimum deformation as the flow reverses its
direction. The stiff capsule therefore feels and responds to the variations of 
the flow fast enough to
change its shape accordingly; following so
closely the flow, its maximum deformation appears very shortly after the 
instant of maximum
shear. The opposite applies to softer capsules.

The temporal evolution of the capsule deformation was also reported in the 
bottom
panel of figure~\ref{fig:nmsts_ulat}. The shaded regions clearly indicate that 
the local maxima of the migration velocity, normal stress difference and 
deformation index appear roughly at a same
moment; moreover, the three quantities vary with twice the frequency of the 
background shear. 
Such a coincidence is not surprising; in fact, for a droplet and a capsule in 
steady shear,
theoretical~\cite{chan1979motion} and numerical~\cite{singh2014lateral} works 
have identified the relation between its 
migration velocity and deformation as $\uld \sim \hD  \left( a /\hid \right) 
^{2}$. Our analysis further confirmed 
that even in the presence of flow oscillation, the deformation and migration 
velocity, which are time-dependent 
{in this case, are correlated.}

\begin{figure}
        \centering
    \includegraphics[width=0.5\columnwidth]
{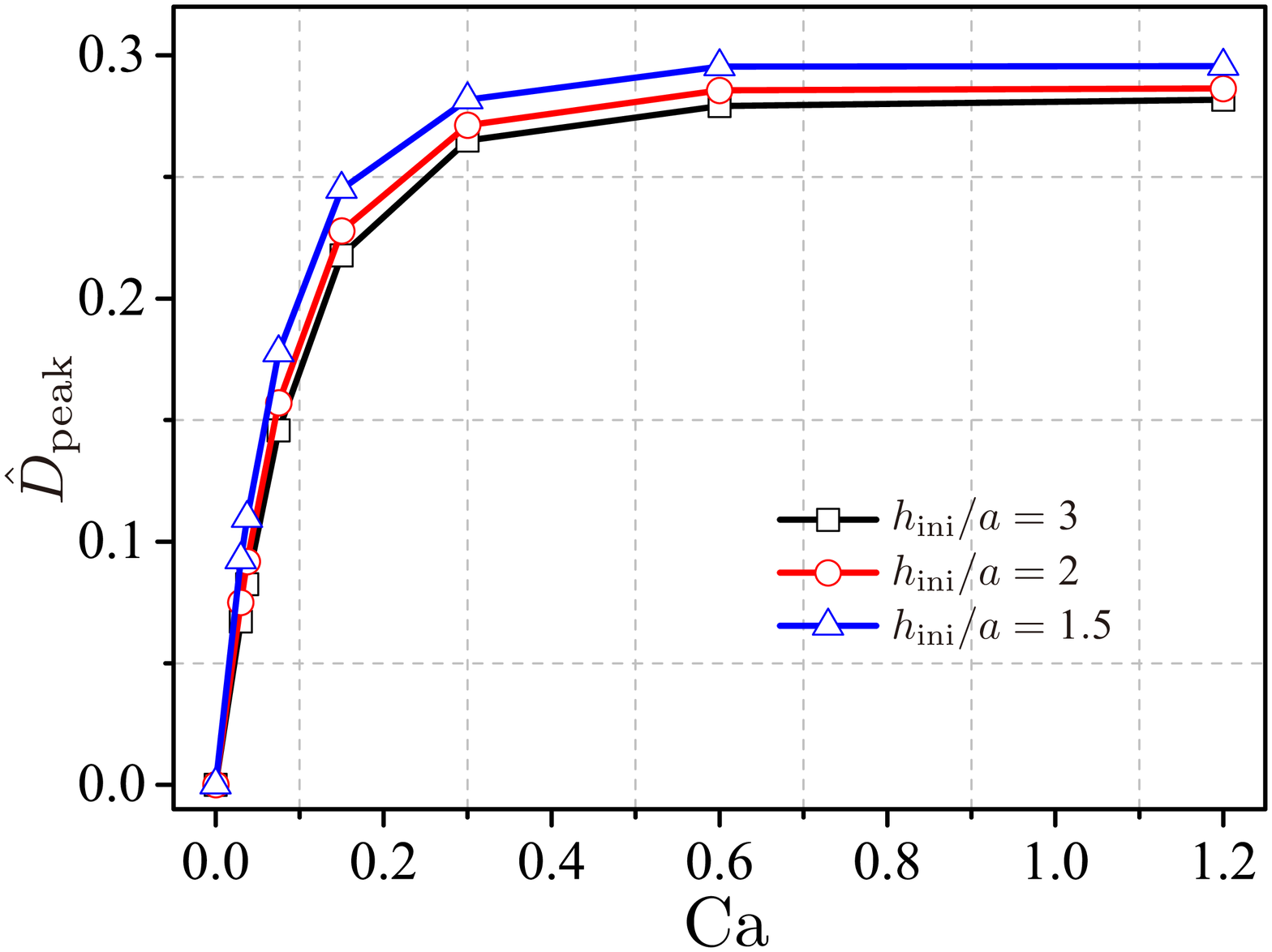}\hspace{0em}       
    \caption{(Color online) The temporal peak of deformation $\hDpk$ versus the
capillary number $\Ca$,  for an initial offset $\hi=1.5$, $2$ and $3$. The 
shear 
frequency is $\omgnon=5/3$.}
\label{fig:dindex_ca_omega1.67}
\end{figure}

To better understand the capsule deformation, we examine the peak deformation 
$\hDpk$, the maximum value of $\hD$ 
after the initial transient overshoot. Figure~\ref{fig:dindex_ca_omega1.67} 
depicts $\hDpk$ as a function of the 
capillary number $\Ca$ for the same three initial heights considered above. 
$\hDpk$ increases almost linearly with the capillary number $\Ca$ for
$\Ca<0.1$. This resembles the linear relation between the Taylor parameter $D$
and the capsule capillary number in an unbounded steady shear flow. Further 
increasing the capillary number,
$\hDpk$ increases more slowly and reaches an asymptotic value close to $0.3$ as
$\Ca>0.6$. This is in contrast to what observed in an unbounded steady shear 
where the deformation $D$ 
always increases with $\Ca$ although slowly at large $\Ca$. 

In oscillating flows, 
the deformation not only depends on $\Ca$ but also on the frequency of the 
oscillations,
$\omgnon$. The time needed to reach the maximum deformation increases with 
$\Ca$; for large values of  $\Ca$, 
the capsule fails to 
reach the maximum possible deformation that would occur in a steady shear 
before the flow changes direction if the 
oscillations occur fast enough. 
Considering $T/2$, the time during which the shear has constant direction, the 
deformation of the capsules
is hence limited by $T$ instead of by its deformability for large values of the 
shear oscillation frequency. 
The relation between the deformation and $\omgnon$ can also explain 
the non-monotonic dependence
of the migration velocity with $\Ca$. 
The lateral migration
velocity, $\ulat$, of  a capsule with $\lambda=1$ in wall-bounded shear
is related to
$S^{yy}_{\mathrm{mem}}$, the $yy$ component of the stresslet 
$\mathbf{S}_{\mathrm{mem}}$ induced by the elastic 
force on the membrane by~\cite{singh2014lateral,nix2014lateral}
\begin{equation}\label{eq:sts_vel}
  \frac{\ulat}{a \dot{\gamma}} = -\frac{9}{64\pi} \left( \frac{a}{\hcw} 
\right)^{2}\frac{S^{yy}_{\mathrm{mem}}}{G_{\mathrm{s}}a^{2}}\frac{1}{\Ca}, 
\end{equation}
where $S^{yy}_{\mathrm{mem}}=\int_{S}-\rho^{\mathrm{e}}_{y}\left( \bx 
\right)\left( 
\bx_{y} - \bx^{\mathrm{cen}}_{y} \right) dS\left(\bx\right)$ and 
$\frac{S^{yy}_{\mathrm{mem}}}{G_{s}a^{2}}$  
 depends only on the capsule  deformation. This 
increases faster at low $\Ca$ and more slowly as $\Ca>0.2$ (see 
figure.~\ref{fig:dindex_ca_omega1.67}) and 
so does $\frac{S^{yy}_{\mathrm{mem}}}{G_{\mathrm{s}}a^{2}}$;
 $\frac{1}{\Ca}$, conversely, decreases so that the product
of the two decreases when $\Ca$ is above a certain value. In another words,
the lateral migration of high-$\Ca$ capsules is hampered by the limiting 
deformation they attain.

\begin{figure}
        \centering
    \subfigure[]{\label{fig:dindex_omega_ca0.3}
    \hspace{-3em}\includegraphics[scale = 0.32] 
{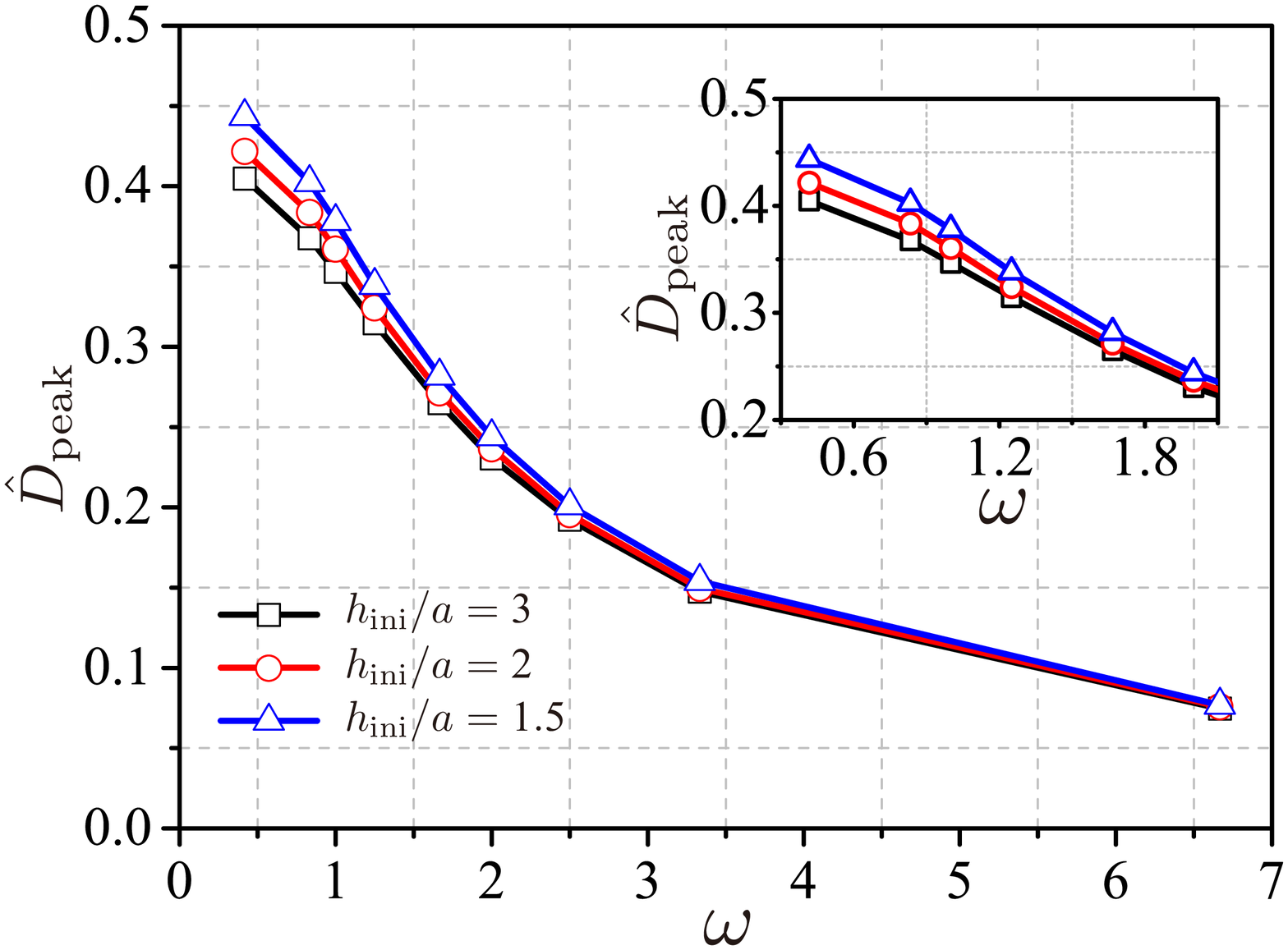}
 }
    \subfigure[]{\label{fig:dindex_omega_ca0.3_scaling}
    \hspace{0em}\includegraphics[scale = 0.32] 
    {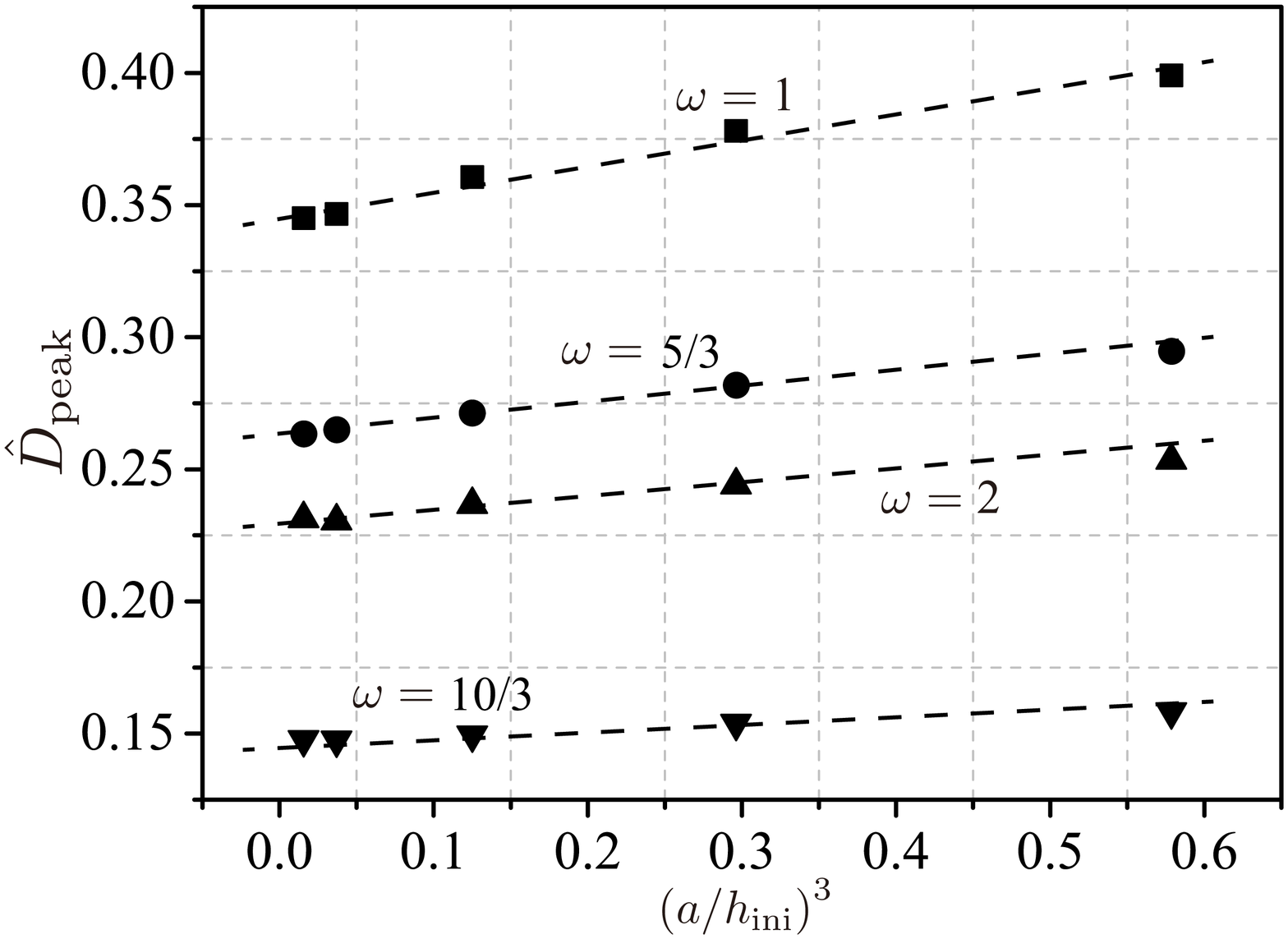}
 }
        \caption{(Color online) \subref{fig:dindex_omega_ca0.3}: the peak 
deformation
        $\hDpk$ as 
a function of the shear frequency $\omgnon$, for $\hi=1.5$, $2$ and $3$ and 
capillary number $\Ca=0.3$. 
\subref{fig:dindex_omega_ca0.3_scaling}: $\hDpk$ versus 
$\left( \adh \right) ^{3}$ for $\omgnon=1$, $5/3$, $2$ and $10/3$.}
\label{fig:dindex_omega_ca0.3_and_scaling}
\end{figure}

We further examine the dependence of $\hDpk$ on the frequency of the shear 
$\omgnon$ and on the initial height 
$\hi$
in figure~\ref{fig:dindex_omega_ca0.3_and_scaling}, where the membrane 
deformability is kept at $\Ca=0.3$. The maximum 
deformation $\hDpk$ decreases 
monotonically with $\omgnon$ as shown in 
figure~\ref{fig:dindex_omega_ca0.3}. $\hDpk$ increases as $\hi$ 
decreases, i.e.\ 
larger confinement effects, 
and this is more evident for smaller values of $\omgnon$; for large 
$\omgnon$, the deformation barely 
varies with 
the initial height for the same mechanism explained above. 
We
show in 
figure~\ref{fig:dindex_omega_ca0.3_scaling} that the maximum deformation 
$\hDpk$ is proportional to 
$\left( \adh \right) ^{3}$. Note that in a wall-bounded steady shear flow, the 
same power law has been identified 
for the deformation of a droplet by Shapira and Haber \cite{shapira1990low} 
theoretically and for a capsule by Singh 
{\it et al.}\cite{singh2014lateral} using numerical simulations.

To quantify the delay between the capsule deformation and the shear 
oscillations, we define the time difference $\lpl 
t_{\mathrm{maxD}}-t_{\mathrm{maxS}} \rp$
scaled by the characteristic flow time as the phase lag $\phi=\lpl
t_{\mathrm{maxD}}-t_{\mathrm{maxS}} \rp/T$. The phase delay $\phi$ is negligible
for the stiff capsule,
whereas it is  $\phi \approx 1/4$ for the floppy capsule with $\Ca=0.6$; the 
latter deforms most when the flow
changes direction (see the time histories in figure 
\ref{fig:dindex_v_time_omega1.67}). 
Note again that, regardless of the difference in the phase delay between 
capsules of different capillary numbers, the 
deformation oscillates at a frequency around $2 \omgnon$.

\begin{figure}
        \centering
    \includegraphics[width=0.5\columnwidth]
{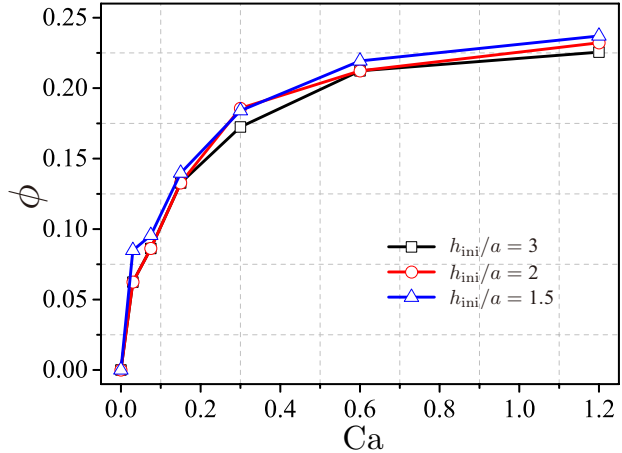}\hspace{0em}       
    \caption{(Color online) Phase delay $\phi$ of the capsule deformation with 
respect to the applied
shear versus the capillary number $\Ca$; the frequency of shear is
$\omgnon=5/3$. Results are shown for an initial offset $\hi=1.5$ 
(triangles), $2$ (circles) and $3$ (squares).}
\label{fig:phasedelay_ca}
\end{figure}

Finally,
we examine in figure~\ref{fig:phasedelay_ca} the dependence of the phase delay 
$\phi$ on the capillary number $\Ca$. The phase delay $\phi$ does not vary 
significantly with 
the initial offset $\hi$ of the capsules. $\phi$ increases monotonically with 
$\Ca$,
an effect more pronounced when  $\Ca$ is below $0.2$.  The delay $\phi$ reaches 
an asymptotic value of approximately
$0.25$ as $\Ca$ continues to increase. Note that the phase delay $\phi$ assumes 
a value between $0$ (for
purely viscous fluid) and $0.5$ (for a linearly elastic solid). The asymptotic 
value of $\approx 0.25$
clearly indicates that viscoelastic effect is an important feature in the flow 
of deformable
cells.

\section{Conclusions}
Using an accelerated boundary integral method for the flow coupled with a global 
spectral method for the membrane,
we study the near-wall dynamics of an initially spherical capsule with 
neo-Hookean membrane in an
oscillating shear flow. We focus on the lateral migration and deformation of the 
capsule, 
and their dependence on the capillary number, the frequency of the background 
shear and the 
initial wall-capsule distance.

The shape of capsule is asymmetric due to the presence of one wall and hence 
its center of mass
and surface centroid are not equivalent, as discussed in 
Ref.~\onlinecite{nix2014lateral}. The time history of the lateral migration of 
the two centroids are clearly different, their time-averaged migrations being 
however similar. It is thus 
important to specify the definition of the centroids of asymmetric deformable 
particles when quantifying their migrations.

The capsule reaches a quasi-periodic steady state after an initial transient period. During this state,
the capsule follows a wiggling trajectory, moving away from and towards the wall 
periodically, with a net lateral migration away from the wall. For the capsule-wall distances and shear 
frequencies investigated here, the mean migration velocity varies non-monotonically with the capillary
number $\Ca$, reaching the maximum migration at an optimal capillary number denoted $\Caopt$.

The optimal capillary number $\Caopt$ is sensitive to the frequency of 
the shear, and is shown to scale linearly with the inverse of $\omgnon$. 
Interestingly, it does not vary
significantly with the capsule-wall distance. The maximum migration is observed when the
effective oscillation period is of the 
order of the relaxation time of the elastic membrane.

The capsule lateral migration velocity decreases monotonically with the frequency of the imposed 
shear; at relatively high oscillation rates, the capsule fails to adapt to the flow before it changes direction so 
that its lateral migration and surface deformation become negligible.

The relation between the lateral migration velocity and the normal stress
difference induced in the flow is explored. The maxima of these two quantities are closely correlated, with a 
weak time delay function of the capillary number. The relative magnitude of the correlations can vary significantly.
As $\Ca \approx \Caopt$, when the capsule has the highest migration 
velocity, the correlation is highest. The dependency of these two quantities
on the capillary number  $\Ca$ is investigated for different capsule-wall distances and oscillation frequencies of the 
imposed shear; in all cases, the two curves agree well with each and the peak values occur as $\Ca \approx \Caopt$. 
This confirms the correlation previously observed  in steady wall-bounded 
shear flows \cite{pranay2012depletion}. It is worth pointing that such a 
relationship has also been discovered for 
a vesicle in  two-dimensional unbounded Couette flow~\cite{ghigliotti2011vesicle}; 
the  vesicle migrates towards the center 
at a velocity linearly scaling with the normal stress difference.

The deformation of the capsule exhibits a periodic variation 
approximately in phase with that of the migration velocity and of the normal stress difference, also at twice the frequency
of the imposed shear. The maximum deformation increases
linearly with $\Ca$ when $\Ca<0.1$ and more slowly as $\Ca>0.2$.
It reaches an asymptotic value when $\Ca$ is above a critical  threshold $\Ca \approx 0.6$; in this regime the 
deformation is limited by the time over which constant shear is applied
and not by the membrane deformability.
The peak deformation is found to scale with the capsule-wall distance $(\hi)^{1/3}$, as 
observed for capsules and droplets in near-wall steady shear~\cite{singh2014lateral}.
We  also discuss the phase delay between the capsule deformation and the 
background shear, 
another viscoelastic feature of the capsule dynamics.

The non-monotonic dependence of the capsule migration velocity 
on its deformability could be potentially used to sort cells. 
Oscillating a suspension 
in a Couette device,
cells with the capillary number exhibiting the largest lateral migration velocity may be extracted and isolated.
By tuning the oscillation frequency, specific cells may therefore be targeted.
The role of the oscillations on the rheological behavior of a suspension is a challenging extension of the present work.

\section*{Acknowledgements}
Computer time provided by SNIC (Swedish National Infrastructure for Computing)
is gratefully acknowledged. We thank the anonymous referees for their 
inspiring comments. Lailai Zhu acknowledges the financial support from the 
European Research Council 
grant (ERC simcomics-280117) as a researcher at EPFL. Jean Rabault acknowledges the financial support received as a 
master student from \'{E}cole Polytechnique. Part of this work was supported by the Linn\'e FLOW
Centre at KTH and by the European Research Council Grant No. 
ERC-2013-CoG-616186, TRITOS to Luca Brandt.

\bibliographystyle{unsrt}
\bibliography{liczhu}

\end{document}